\documentclass[notitlepage,twocolumn,letterpaper,natbib,aps,prd,amsmath,amsfonts,nofootinbib,preprintnumbers,superscriptaddress,secnumarabic%
]{revtex4-1}
\pdfoutput=1
\usepackage{amssymb,amsmath,latexsym,mathrsfs}
\usepackage{url}
\usepackage{enumitem}
\usepackage{graphicx}
\usepackage{booktabs}
\usepackage[usenames,dvipsnames]{color}
\usepackage[breaklinks,colorlinks,urlcolor=magenta,citecolor=magenta,linkcolor=magenta]{hyperref}
\usepackage{multirow}
\usepackage{float}
\usepackage{cases}
\usepackage{blindtext}
\usepackage{pifont}
\usepackage{hhline}
\usepackage[normalem]{ulem}

\usepackage{xcolor}
\definecolor{linkcolor}{HTML}{FF6461}
\definecolor{citecolor}{rgb}{1.0, 0.5, 0.0}
\definecolor{highpcol}{HTML}{3F7BB6}
\definecolor{lowtcol}{HTML}{800080}
\definecolor{lcdmcol}{HTML}{B8860B}
\hypersetup{
  linkcolor  = linkcolor,
  citecolor  = linkcolor,
  urlcolor   = linkcolor,
  colorlinks = true
}

\begin{document}

\title{Cosmic Neutrino Background Detection in Large-Neutrino-Mass Cosmologies}

\author{James Alvey}
\email{j.b.g.alvey@uva.nl}
\thanks{ORCID: \href{https://orcid.org/0000-0003-2020-0803}{0000-0003-2020-0803}}
\affiliation{GRAPPA Institute, Institute for Theoretical Physics Amsterdam,\\
University of Amsterdam, Science Park 904, 1098 XH Amsterdam, The Netherlands}

\author{Miguel Escudero}
\email{miguel.escudero@tum.de}
\thanks{ORCID: \href{https://orcid.org/0000-0002-4487-8742}{0000-0002-4487-8742}}
\affiliation{Physik-Department, Technische Universit{\"{a}}t, M{\"{u}}nchen, James-Franck-Stra{\ss}e, 85748 Garching, Germany}

\author{Nashwan Sabti}
\email{nashwan.sabti@kcl.ac.uk}
\thanks{ORCID: \href{https://orcid.org/0000-0002-7924-546X}{0000-0002-7924-546X}}
\affiliation{Department of Physics, King's College London, Strand, London WC2R 2LS, UK}

\author{Thomas Schwetz}
\email{schwetz@kit.edu}
\thanks{ORCID: \href{https://orcid.org/0000-0001-7091-1764}{0000-0001-7091-1764}}
\affiliation{Institut f\"ur Astroteilchenphysik, Karlsruher Institut f\"ur Technologie (KIT), Hermann-von-Helmholtz-Platz 1, 76344 Eggenstein-Leopoldshafen, Germany}

\preprint{TUM-HEP-1374/21,  KCL-2021-88}

\begin{abstract}
 
\noindent The Cosmic Neutrino Background (CNB) is a definite prediction of the standard cosmological model and its direct discovery would represent a milestone in cosmology and neutrino physics. In this work, we consider the capture of relic neutrinos on a tritium target as a possible way to detect the CNB, as aimed for by the PTOLEMY project. Crucial parameters for this measurement are the absolute neutrino mass $m_\nu$ and the local neutrino number density $n_\nu^{\rm loc}$. Within the $\Lambda$CDM model, cosmology provides a stringent upper limit on the sum of neutrino masses of $\sum m_\nu < 0.12\,{\rm eV}$, with further improvements expected soon from galaxy surveys by DESI and EUCLID. This makes the prospects for a CNB detection and a neutrino mass measurement in the laboratory very difficult. In this context, we consider a set of non-standard cosmological models that allow for large neutrino masses ($m_\nu \sim 1\,{\rm eV}$), potentially in reach of the KATRIN neutrino mass experiment or upcoming neutrinoless double-beta decay searches. We show that the CNB detection prospects could be much higher in some of these models compared to those in $\Lambda$CDM, and discuss the potential for such a detection to discriminate between cosmological scenarios. Moreover, we provide a simple rule to estimate the required values of energy resolution, exposure, and background rate for a PTOLEMY-like experiment to cover a certain region in the $(m_\nu,\, n_\nu^{\rm loc})$ parameter space. Alongside this paper, we publicly release a code to calculate the CNB sensitivity in a given cosmological model.

\vspace*{5pt} \noindent \textbf{\texttt{GitHub}}: Public code to compute the sensitivity of a PTOLEMY-like experiment can be found \href{https://github.com/james-alvey-42/DistNuAndPtolemy}{here}.

\end{abstract}

\maketitle
\hypersetup{
  linkcolor  = linkcolor,
  citecolor  = linkcolor,
  urlcolor   = linkcolor,
  colorlinks = true
}

\section{Introduction}

\subsection*{The Cosmic Relic Neutrino Background}

\vspace{-0.3cm}

\noindent One of the central predictions of the standard cosmological model is the existence of the Cosmic Relic Neutrino Background (CNB). In the $\Lambda$CDM scenario, we expect a neutrino population with a momentum distribution close to the thermal Fermi-Dirac distribution
\cite{Dolgov:1997mb,Mangano:2005cc,deSalas:2016ztq,Escudero:2018mvt}, 
with a present day temperature of:
\begin{equation}\label{eq:Tnu}
 T_{\nu,0}^{\rm SM} \approx T_{\gamma,0}/1.4 \approx 1.95 \, {\rm K}\ ,
\end{equation}
and an average number density of about:
\begin{equation}\label{eq:n_nu0}
  n_{\nu,0}^{\rm SM} = \frac{3}{4} \frac{\zeta(3)}{\pi^2}\, T_{\nu,0}^3 
  \approx 56\,{\rm cm}^{-3}\ ,
\end{equation}
for each helicity degree of freedom. Its existence has been
established indirectly at very high confidence by the determination of
the effective number of relativistic species in the early Universe, $N_{\rm eff}$, both via measurements of the primordial element abundances as synthesised during Big Bang
Nucleosynthesis (BBN), as well as by observations of the Cosmic
Microwave Background (CMB). A recent global BBN analysis~\cite{Pisanti:2020efz} (see also~\cite{Pitrou:2018cgg,Fields:2019pfx}) obtains:
\begin{equation}\label{eq:NeffBBN}
  N_{\rm eff} = 2.78\pm 0.28 \,(68\%\, {\rm CL})\ ,
\end{equation}
when using the latest helium primordial abundance from~\cite{Hsyu:2020uqb}, the deuterium measurements from~\cite{Cooke:2017cwo}, and an updated set of nuclear reaction rates from \cite{Mossa:2020gjc}. From CMB observations, combined with Baryonic Acoustic Oscillations (BAO), the Planck collaboration reports \cite{Aghanim:2018eyx}:
\begin{equation}\label{eq:NeffCMB}
  N_{\rm eff} = 2.99\pm 0.17 \,(68\%\, {\rm CL}) \ .
\end{equation}
Both of these numbers are in remarkable agreement with each other, as
well as with the prediction of the standard $\Lambda$CDM model of
$N_{\rm eff} = 3.044$~\cite{EscuderoAbenza:2020cmq,Akita:2020szl,Bennett:2020zkv,Froustey:2020mcq}. Moreover, they are different from zero at very high confidence, implying an indirect detection of the presence of cosmological neutrinos.

The direct detection of relic neutrinos by experiments on Earth, however, is very challenging. This is mainly because their interaction cross-section is tiny as a result of the very low neutrino energies, see Eq.~\eqref{eq:Tnu}. A possible method to overcome the low energy deposition is to use the capture on a $\beta$-unstable nucleus, which is a threshold-less reaction \cite{Weinberg:1962zza,Cocco:2007za}. PTOLEMY~\cite{PTOLEMY:2018jst} is an ambitious project pursuing this idea using electron neutrino capture on tritium:
\begin{equation}\label{eq:beta_capture}
  \nu_e + {}^3{\rm H} \to e^- + {}^3{\rm He}^{+} \ .
\end{equation}
The signature of the relic neutrino background would be a peak in the electron energy above the continuous beta-spectrum endpoint, separated from the endpoint by about twice the
neutrino mass. Experimentally key quantities in this regard are the amount of available
tritium for the target, very low backgrounds, as well as the excellent
energy resolution of the detector needed to separate the
CNB-induced peak from the $\beta$-decay continuum\footnote{Indeed, there could be even some fundamental physics limitations to achieve the required resolution \cite{Cheipesh:2021fmg,Nussinov:2021zrj}.}.
Furthermore, the detection of the signal becomes exceedingly difficult the smaller the neutrino mass is, since this controls the separation of the CNB-induced peak from the $\beta$-decay background. For phenomenological studies and sensitivity estimates, see e.g., \cite{Blennow:2008fh,Long:2014zva,PTOLEMY:2019hkd,Akita:2020jbo}. 
In spite of these challenges, the detection of the CNB would be an outstanding experimental achievement and provide a window into the very early stages of the Universe.

\subsection*{Neutrino Masses: \\ 
Current Status and Future Prospects}

\vspace{-0.3cm}

\noindent From the particle physics side, an important open question in neutrino
physics is the absolute mass scale of neutrinos~\cite{Formaggio:2021nfz}. The mass-squared differences of the three neutrino mass states, $\Delta m^2_{ji} \equiv
m_j^2-m_i^2$ ($i,j=1,2,3$), are determined by oscillation experiments
with few percent precision~\cite{Esteban:2020cvm} (see also~\cite{deSalas:2020pgw,Capozzi:2021fjo}):
\begin{align}
\label{eq:mass_splittings_1}
\Delta m^2_{21} &=
(7.42\pm0.21)\times 10^{-5}\,\mathrm{eV}^2\ ,\\
\label{eq:mass_splittings_2}
\Delta m^2_{31} &=
(2.514\pm0.028)\times 10^{-3}\,\mathrm{eV}^2\ \mathrm{or}\nonumber\\ \Delta m^2_{32} &=
-(2.497\pm0.028)\times 10^{-3}\,\mathrm{eV}^2\ ,    
\end{align}
where the two possible choices
for the larger mass-squared difference correspond to the so-called
normal ordering (NO) and inverted ordering (IO) of the mass states,
respectively.  Once the mass-squared differences are fixed, the
absolute mass scale can be parameterised by the mass of the lightest
neutrino, $m_{\rm lightest}$, where in the standard convention $m_{\rm
  lightest} = m_1\,(m_3)$ for NO (IO).

From the observation of the beta-decay spectrum close to the endpoint,
one can constrain the effective neutrino mass:
\begin{equation}
  m_\beta^2 = \sum_{i=1}^3 m_i^2|U_{ei}|^2 \ ,
\end{equation}
where $U_{ei}$ are the leptonic mixing matrix elements, whose moduli are
determined with good precision by oscillation experiments, see
e.g.~\cite{Esteban:2020cvm}. The current best limit on $m_\beta$ comes from the KATRIN experiment and is given by~\cite{KATRIN:2019yun,Aker:2021gma}:
\begin{equation}\label{eq:Katrin_bound}
  m_\beta < 0.8\,{\rm eV}\,(90\%\,{\rm CL}) \ ,
\end{equation}
while the final sensitivity goal of KATRIN is $0.2\,\mathrm{eV}$, something that will be reached within a few years. For values of $m_\beta$ in this regime, neutrinos
are quasi-degenerate and $m_\beta \approx m_{\rm lightest}$. Besides KATRIN, there are other projects aiming to measure the neutrino mass, including Project 8~\cite{Project8:2017nal} and ECHo~\cite{Gastaldo:2017edk}.

On the other hand, within the $\Lambda$CDM model, cosmology provides a tight bound on the sum of neutrino masses due their impact on cosmological structure formation. From combined CMB and BAO observations, the Planck collaboration obtains~\cite{Aghanim:2018eyx}:
\begin{equation}\label{eq:sum_bound}
  \sum m_\nu \equiv \sum_{i=1}^3m_i < 0.12\,{\rm eV} \,(95\%\,{\rm CL}) \ ,
\end{equation}
which, when taken at face value, implies:
\begin{equation}\label{eq:m0_planck}
  m_{\rm lightest} <
  \left\{
  \begin{array}{l}
    0.03\,{\rm eV \quad(NO)} \\
    0.016\,{\rm eV \quad(IO)}
  \end{array}\right.  \,.
\end{equation}
Depending on the precise cosmological data used, in principle even stronger limits can be obtained, see
e.g.~\cite{DiValentino:2021hoh}. Indeed, we can expect that with data
from future large-scale structure surveys by
DESI~\cite{DESI:2016fyo} and Euclid~\cite{Amendola:2016saw},
sensitivities to $\sum m_\nu$ of $0.02\,\mathrm{eV}$ could be achieved, see e.g.
\cite{Brinckmann:2018owf}. Note that from neutrino oscillation data, a
minimal value of $\sum m_\nu \approx 0.06\,\mathrm{eV}$ for NO and $0.1\,\mathrm{eV}$ for IO
is predicted for $m_{\rm lightest} = 0$. Hence, we could expect a
positive detection of a finite neutrino mass from cosmology in $\Lambda$CDM soon.

If neutrinos are Majorana particles, they will induce the
lepton-number violating process neutrinoless double-beta decay. In the
absence of cancellations due to other exotic physics, the corresponding
decay rate can be related to an effective Majorana mass:
\begin{equation}\label{eq:mbb}
  m_{\beta\beta} = \left|\sum_{i=1}^3 m_i U_{ei}^2\right| \ .
\end{equation}
Note that this relation depends on unknown complex phases of $U_{ei}$,
the so-called Majorana phases. Current strongest constraints come
from the KamLAND-Zen experiment~\cite{KamLAND-Zen:2016pfg}, leading to:
\begin{equation}\label{eq:mbb_bound}
  \begin{array}{rl}
  m_{\beta\beta} &< 0.061 - 0.165 \, {\rm eV} \\    
  m_{\rm lightest} &< 0.180 - 0.480 \, {\rm eV} 
  \end{array}
 \quad (90\% \, {\rm CL}) \ ,
\end{equation}
where the indicated range corresponds to the uncertainty from nuclear
matrix elements, and for the limit on $m_{\rm lightest}$ the
least-constraining values of the Majorana phases have been adopted.
For other recent results with comparable sensitivity, see
\cite{GERDA:2020xhi,CUORE:2021gpk,EXO-200:2019rkq}.
Furthermore, there is strong ongoing experimental effort to reach sensitivities in the range
$m_{\beta\beta} \approx 0.01 - 0.02$~eV~\cite{Giuliani:2019uno}.

\subsection*{Large Neutrino Mass Cosmologies and the CNB}
\vspace{-0.3cm}

\noindent Comparing the cosmological bound from Eq.~\eqref{eq:m0_planck} with the KATRIN sensitivity, we see that in the standard scenario, the neutrino mass should be out of reach for KATRIN. Moreover, also in view of the cosmological bound on the neutrino mass, a direct detection of the CNB with the PTOLEMY project seems improbable, even under optimistic assumptions~\cite{PTOLEMY:2019hkd}. However, the strong link between the quantities constrained by cosmology and the terrestrial experiments today relies both on standard cosmology and particle physics. As shown in our companion paper~\cite{Alvey:2021sji}, if the neutrino distribution is free, then current CMB(+BAO) data is actually not able to measure neutrino masses directly, but rather only the non-relativistic neutrino energy density:
\begin{align}
    \label{eq:rhoNR_bound}
    \rho_{\nu,0}^\mathrm{NR}  = \sum m_\nu (2\,n_{\nu,0}) < 14\,\text{eV}\text{cm}^{-3}\ ,
\end{align}
which is a product of the neutrino mass and number density. Thus, by sufficiently reducing the number density of neutrinos, the mass bound may be weakened. Indeed, several mechanisms are known in the literature which would relax the cosmological neutrino mass bound and allow for a neutrino mass that would be observable in KATRIN as well as in neutrinoless double-beta decay experiments (in the case of Majorana neutrinos). Such scenarios are fully consistent with cosmological data and include neutrino decays~\cite{Chacko:2019nej,Escudero:2020ped,Escudero:2019gfk,Barenboim:2020vrr,Chacko:2020hmh,Abellan:2021rfq}, neutrinos with a time-varying mass~\cite{Dvali:2016uhn,Dvali:2021uvk,Lorenz:2018fzb,Lorenz:2021alz,Esteban:2021ozz}, neutrinos with a temperature much lower than the thermal one in Eq.~\eqref{eq:Tnu} supplemented with dark radiation~\cite{Farzan:2015pca,GAMBITCosmologyWorkgroup:2020htv}, and neutrinos with a distribution function that deviates from the Fermi-Dirac one~\cite{Cuoco:2005qr,Oldengott:2019lke,Alvey:2021sji}. 

In this paper, we aim to we investigate such ``large-neutrino-mass cosmologies'' within the context of a CNB detection. While we will review each of the above cosmological settings in Sec.~\ref{sec:scenarios}, in this work we will focus on the detection prospects of the scenarios with non-standard neutrino populations. Namely, those that have a temperature $T_\nu < T_\nu^{\rm SM}$ as well as a dark radiation component, and those with significant deviations from the usual Fermi-Dirac distribution. We choose these two in particular, because the detection prospects of the other scenarios at PTOLEMY are either null, require extremely optimistic experimental configurations, or lead to unpredictable rates at PTOLEMY. For instance, in the presence of neutrino decays~\cite{Chacko:2019nej,Escudero:2020ped,Escudero:2019gfk,Barenboim:2020vrr,Chacko:2020hmh,Abellan:2021rfq}, one expects no relic active neutrino background today if the neutrino mass is larger than $m_\nu \gtrsim 0.1\,\text{eV}$. On the other hand, if the neutrino mass is small enough, then it is possible that there still exist relic neutrinos today. However, a CNB detection in this case would require a highly optimistic configuration of PTOLEMY\footnote{If such an optimistic configuration can be achieved, then a number of interesting conclusions about the neutrino mass and lifetime could be inferred, see~\cite{Akita:2021hqn}.}. Another example comes from scenarios involving neutrinos with a time-dependent mass. Some of these setups lead to the absence of any cosmic neutrinos at the present time, see e.g.~\cite{Dvali:2016uhn,Dvali:2021uvk}, whilst in other cases, one does expect a relic population today~\cite{Esteban:2021ozz}. In the latter case, however, the actual number density of neutrinos in the Milky Way cannot be reliably predicted, as it is subject to a non-linear cosmological evolution that is not yet fully understood~\cite{Ayaita:2014una,Casas:2016duf}. Regardless of these challenges, the interplay of possible effects at PTOLEMY make it a key experiment potentially capable of differentiating and distinguishing between the various cosmological settings.

\subsection*{Goals and Structure of this Study} 

\noindent Our aim with this study is to understand the possibilities of a Cosmic Neutrino Background detection within the context of large neutrino mass cosmologies at a PTOLEMY-like experiment. We believe that this is an interesting topic on its own, as the detection of the CNB would be a significant milestone for cosmology and particle physics, but we are also motivated by a number of other important factors: \textit{i)} the current experimental efforts to measure $m_\nu$ and $m_{\beta \beta}$ in the laboratory, \textit{ii)} the very stringent cosmological constraints on the neutrino mass within $\Lambda$CDM, \textit{iii)} the prospects to potentially detect the neutrino mass with ongoing/upcoming galaxy surveys, and \textit{iv)} the theoretical landscape of models where neutrinos could have a large mass. 

We then structure this paper as follows: Firstly, in Sec.~\ref{sec:cosmopheno} we discuss the main cosmological features controlling the detection prospects of the CNB. In particular, we discuss and calculate the allowed ranges for $m_\nu$ and the local number density of neutrinos within $\Lambda$CDM and the two main non-standard cosmologies under consideration. Next, in Sec.~\ref{sec:pheno}, we outline our approach in obtaining the sensitivity of a PTOLEMY-like experiment to detect the CNB. In Sec.~\ref{sec:results}, we present our results for the expected sensitivity. In particular, we discuss the CNB detection sensitivity as a function of the various experimental factors, and examine it within the context of $\Lambda$CDM and the two non-standard cosmologies. We also consider future experimental situations in light of expected data from KATRIN, DESI/EUCLID and the next generation of neutrinoless double-beta decay experiments. In Sec.~\ref{sec:scenarios}, we review other non-standard cosmological scenarios where neutrinos can have a large mass and comment on how a CNB search at PTOLEMY could distinguish between them. Finally, in Sec.~\ref{sec:conclusions} we summarise our results and present our conclusions. Supplementary details regarding the clustering of neutrinos in the Milky Way are provided in App.~\ref{app:nu_clustering}.

\begin{figure*}[t!]
    \centering
    \includegraphics[width=0.98\linewidth]{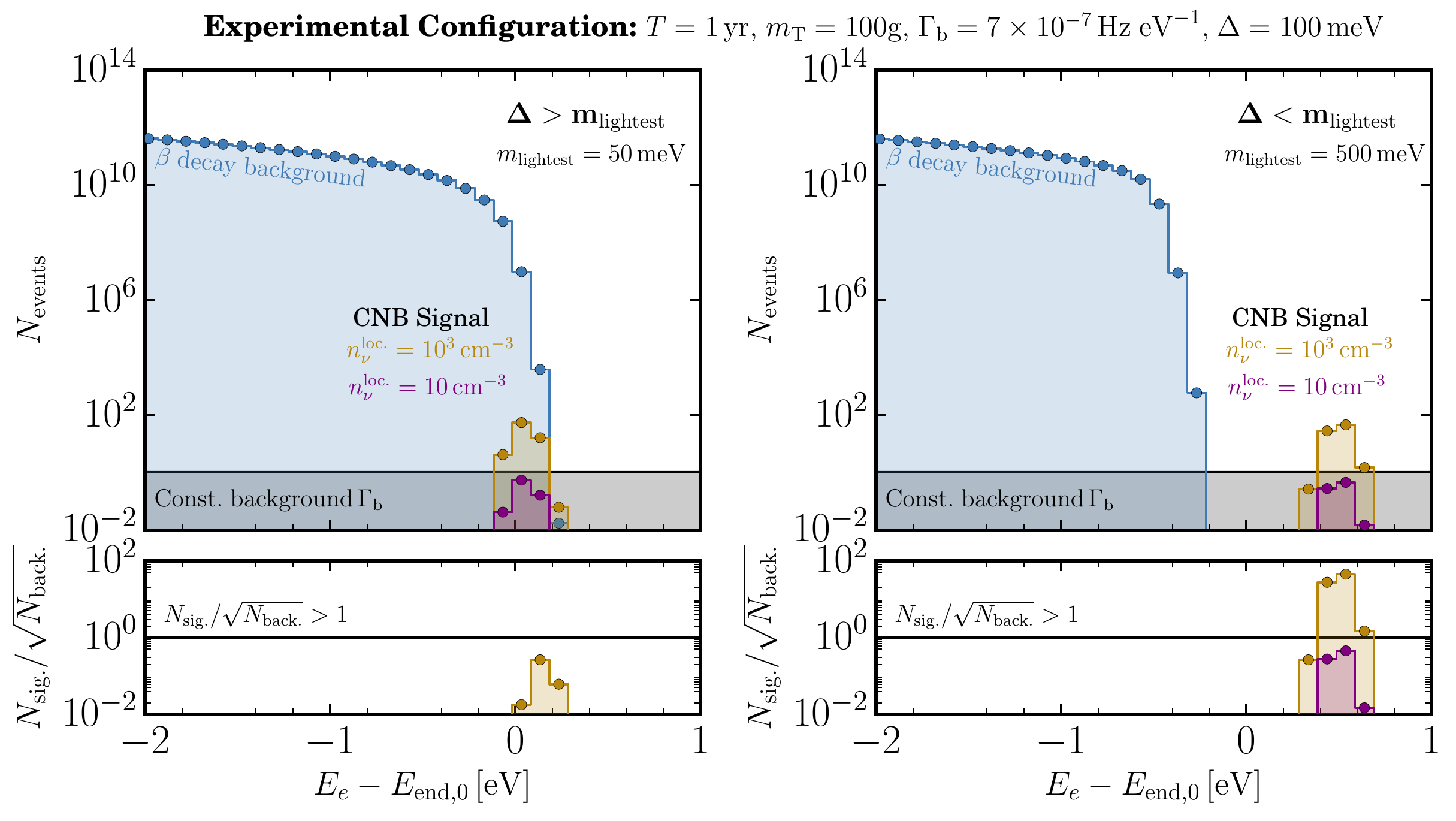}
    \caption{Event spectra expected at PTOLEMY within the fiducial scenario considered in this work for the two cases where the energy resolution is $\Delta > m_{\rm lightest}$ (\emph{left}) and $\Delta < m_{\rm lightest}$ (\emph{right}), see Sec.~\ref{sec:pheno} for details. Note that the sensitivity of a PTOLEMY-like experiment to detect the Cosmic Neutrino Background is governed by both \emph{a}) the separation of the CNB signal from the large $\beta$-decay background, and \emph{b}) the absolute CNB signal rate, specifically as compared to the background. The former is controlled by the relative sizes of $\Delta$ and $m_\mathrm{lightest}$, while the latter is instead specified by the exposure, local number density $n_\nu^\mathrm{loc}$, and the background rate $\Gamma_\mathrm{b}$.}
    \label{fig:spectrum}
\end{figure*}

\section{CNB detection at PTOLEMY: Cosmological Models}
\label{sec:cosmopheno}

\noindent The detection prospects of the CNB via inverse $\beta$-decay capture in tritium, $\nu_e + {}^3{\rm H} \to e^- + {}^3{\rm He}^{+}$, depend upon several experimental factors which we discuss in detail below. However, in terms of the cosmological model, there are only two relevant features that are important: \textit{i)} the neutrino mass $m_\nu$, and $\textit{ii)}$ the number density of relic neutrinos on Earth $n_\nu^{\rm loc}$. The neutrino mass is critical since it controls the minimum energy resolution that needs to be achieved for a successful detection of the CNB. This is because there is a huge continuous $\beta$-decay background from ${}^3{\rm H} \to e^- + {}^3{\rm He}^{+} + \bar{\nu}_e $ decays that is only separated by $\Delta E_e \sim 2 m_\nu$ from the small cosmological signal arising from neutrino capture, see Fig.~\ref{fig:spectrum}. This means that a CNB detection requires a very small energy resolution of at least $\Delta \lesssim m_\nu$. In addition, of course, the number density of relic neutrinos on Earth is critical for the detection because it essentially controls the rate at which the $\nu_e + {}^3{\rm H} \to e^- + {}^3{\rm He}^{+}$ process occurs, given that $\Gamma_\mathrm{CNB} \propto n_{\nu}^{\rm loc}$. Clearly, the larger the number density of neutrinos, the better the sensitivity to detect the CNB will be, see again Fig.~\ref{fig:spectrum}.

Now, since the current cosmological neutrino mass bound from Planck is very stringent within $\Lambda$CDM (see Eq.~\eqref{eq:sum_bound}), and achieving such a correspondingly small value of $\Delta$ is technologically very challenging~\cite{PTOLEMY:2018jst}, this suggests that prospects for detection are likely more reasonable in non-standard cosmologies with large neutrino masses. As mentioned in the introduction, in this regard, it is worth emphasising that the CMB is not directly sensitive to the neutrino mass, but rather to the non-relativistic neutrino energy density $\rho_{\nu,0}^\mathrm{NR}$~\cite{Alvey:2021sji}. This is the product of the neutrino mass and the cosmological neutrino number density per species $(2n_{\nu,0})$, and the current Planck+BAO bound on this quantity is given in Eq.~\eqref{eq:rhoNR_bound}.
From this observation, we can clearly appreciate that a simple way to substantially relax current cosmological bounds on neutrino masses is to reduce the cosmological neutrino number density. Assuming that $n_{\nu,0}$ is the same for each neutrino mass state, this corresponds to a maximum neutrino number density per helicity state of:
\begin{align}
    n_{\nu,0}  < 56\,\text{cm}^{-3} \frac{0.12\,\text{eV}}{\sum m_\nu}\ .
\end{align}
From a phenomenological perspective, we consider two examples in which the neutrino number density in the early Universe is reduced, while being in agreement with all known cosmological measurements:\\

\noindent {\color{lowtcol}{\textbf{Low-}\boldmath{$T_\nu$}\textbf{+DR}}}: \textit{Neutrinos with a temperature $T_\nu < T_\nu^{\rm SM}$ and dark radiation.} The first possibility is that neutrinos have a temperature smaller than the one expected within the Standard Model~\cite{Farzan:2015pca,GAMBITCosmologyWorkgroup:2020htv}. Since $n_\nu \propto T_\nu^3$, this means that even with a slightly lower temperature than in $\Lambda$CDM (as in Eq.~\eqref{eq:Tnu}), the number density in neutrinos can be substantially smaller. Therefore, the neutrino mass can be larger while still satisfying Planck CMB constraints. In particular, the neutrino mass bound can be relaxed as:
\begin{align}\label{eq:mnu_Tnudiff}
    \sum m_\nu < 0.12\,\text{eV} \, \left[\frac{T_\nu^{\rm SM}}{T_\nu}\right]^3\ .
\end{align}
However, reducing the number density of neutrinos in this manner will also impact the energy density in ultrarelativistic species, because $N_{\rm eff}^\nu \propto T_\nu^4$. Therefore, for this setting to be fully compatible with CMB data, one needs to introduce a certain amount of dark radiation to compensate for the decrease of $N_{\rm eff}^\nu$ from neutrinos, such that the total $N_{\rm eff} = N_\mathrm{eff}^\nu + N_\mathrm{eff}^\mathrm{DR} \sim 3$. Indeed, there already exists a proposed mechanism that could achieve this, through the addition of new massless states beyond the Standard Model, see~\cite{Farzan:2015pca}.\\ 

\noindent {\color{highpcol}{\textbf{High-}\boldmath{$p_\nu$}}}: \textit{Neutrinos with an average momentum $\left<p_\nu\right> > 3.15\,T_\nu^{\rm SM} $}. This second possibility considers neutrinos with an average momentum larger than the one in $\Lambda$CDM~\cite{Alvey:2021sji}, see also~\cite{Oldengott:2019lke}. Since $N_{\rm eff}^\nu \propto \rho_\nu|_{p_\nu \gg m_\nu} = n_\nu \left<p_\nu\right> \sim 3$, then in this scenario no dark radiation needs to be introduced and the neutrino mass bound can be relaxed as: 
\begin{align}\label{eq:mnu_pggTnu}
    \sum m_\nu < 0.12\,\text{eV} \,\frac{\left<p_\nu\right>}{3.15\,T_\nu^{\rm SM}}\ ,
\end{align}
where $3.15\,T_\nu^{\rm SM}$ is the average momentum of neutrinos in $\Lambda$CDM and we have assumed $N_\mathrm{eff}^\nu = 3.044$. We see that the larger $\left<p_\nu\right>$ is, the more the neutrino mass bound can be relaxed. It is important to note that so far there is no known mechanism capable of significantly increasing $\left<p_\nu\right>$~\cite{Alvey:2021sji}, but that there are particle physics scenarios (such as sterile neutrino decays during the BBN epoch~\cite{Sabti:2020yrt}) that could potentially lead to $\left<p_\nu\right> \gtrsim 3.15\,T_\nu^\mathrm{SM}$ and weaken the neutrino mass constraint appreciably, see~\cite{Oldengott:2019lke}.

\subsection{Neutrino Number Density on Earth}
\noindent Importantly for CNB detection, the quantity that determines the rate of neutrino capture events is not the cosmological number density $n_{\nu, 0}$, but instead the number density of relic neutrinos on Earth, $n_{\nu}^{\rm loc}$. These two numbers are expected to be different, essentially because the Milky Way gravitational potential causes a clustering of neutrinos in its halo~\cite{Ringwald:2004np}, see also~\cite{deSalas:2017wtt,Zhang:2017ljh,Mertsch:2019qjv}. The main proxy to understand how many neutrinos can cluster is their velocity $v_\nu$. This is because neutrinos with a velocity larger than the escape velocity of the Milky Way cannot become gravitationally bound. In the three cosmological settings discussed above, and assuming degenerate neutrinos that are non-relativistic, we have:
\begin{flalign}
 v_{\nu}(z)  &= 4000  (1\!+\!z) \left[\frac{0.04\,{\rm eV}}{m_\nu}\right] \,\mathrm{km\,s}^{-1} \ \ \,\,\,\,\,\, [{\color{lcdmcol}{\Lambda {\rm CDM}}}], \label{eq:vLCDM}\\
 v_{\nu}(z) &= 4000\, (1\!+\!z)   \left[\frac{0.04\,{\rm eV}}{m_\nu}\right]^{4/3} \,\mathrm{km\,s}^{-1} \, [{\color{lowtcol}{\mathbf{Low}\text{-}\boldsymbol{T_\nu}\mathbf{+DR}}}]\,, \label{eq:vsmallT}\\
 v_{\nu}(z) & = 4000\, (1\!+\!z) \ \, \mathrm{km\,s}^{-1} \,\,    \qquad\qquad \quad \,\,\, [{\color{highpcol}{\mathbf{High}\text{-}\boldsymbol{p_\nu}}}],\hspace*{-10pt} \label{eq:vhighp}
\end{flalign}
where the mean velocity for non-relativistic particles is simply given by $v_\nu = \left<p_\nu\right>/ m_\nu $. For the $\Lambda{\rm CDM}$ case we have $\left<p_\nu\right>^{\rm SM} = 3.15\,T_\nu^{\rm SM}$. For the Low-$T_\nu$+DR scenario, the mean momentum is given by $\left<p_\nu\right>= 3.15\,T_\nu$, and to find Eq.~\eqref{eq:vsmallT} we use the $T_\nu$ that saturates the bound in Eq.~\eqref{eq:mnu_Tnudiff}. Finally, for the High-$p_\nu$ case, we use the mean neutrino momentum that saturates the bound in Eq.~\eqref{eq:mnu_pggTnu}, which explains the absence of a neutrino mass dependence in Eq.~\eqref{eq:vhighp}. Note that Eqs.~\eqref{eq:vLCDM}$-$\eqref{eq:vhighp} only hold for redshifts $z$ where $v_\nu(z) \ll c$.

The escape velocity of the Milky-Way galaxy is roughly ${\sim}\,550\,\text{km\,s}^{-1}$~\cite{Piffl:2013mla}. Comparing the numbers above with this one, we can see that in all of these scenarios we do not expect a substantial gravitational clustering of neutrinos if their masses are smaller than $m_\nu \lesssim 0.3\,\text{eV}$. In fact, in the High-$p_\nu$ cosmological setting, the velocity of neutrinos is so large that independently of the neutrino mass, the clustering should be insignificant. On the other hand, in the Low-$T_\nu$+DR cosmology, one can expect substantial clustering for large neutrino masses. Note also that in $\Lambda$CDM the sum of neutrino masses is bounded to be $\sum m_\nu < 0.12\,\text{eV}$, and so there should not be any significant clustering for masses that are cosmologically allowed. 

In order to accurately model the gravitational clustering of neutrinos in the Milky Way, one should resort to N-body simulations that track the evolution of the distribution function of neutrinos in the evolving gravitational potential, as done in~\cite{Ringwald:2004np,deSalas:2017wtt,Zhang:2017ljh,Mertsch:2019qjv} within $\Lambda$CDM. As an approximation, however, Ref.~\cite{Ringwald:2004np} proposed a method based on linear perturbation theory that allows one to find an estimate of the neutrino clustering for any primordial neutrino distribution function. We discuss the approach in App.~\ref{app:nu_clustering} and use it to compute the clustering for the two non-standard cosmologies we consider. For the $\Lambda$CDM cosmology, we resort to the N-body result of~\cite{Zhang:2017ljh}. In short, the clustering factors read:
\begin{align}
 f_\mathrm{c}  &\simeq 77 \left({m_\nu}/{\text{eV}}\right)^{2.2}    &[{\color{lcdmcol}{\Lambda {\rm CDM}}}]\,, \\
 f_\mathrm{c} & \simeq 96\left({m_\nu}/{\text{eV}}\right)^{2.0}        &[{\color{lowtcol}{\mathbf{Low}\text{-}\boldsymbol{T_\nu}\mathbf{+DR}}}]\,, \label{eq:fc_Tnusmall}\\
 f_\mathrm{c} & \simeq 0     &[{\color{highpcol}{\mathbf{High}\text{-}\boldsymbol{p_\nu}}}]\,,\label{eq:fc_pnularge}
\end{align}
where, following~\cite{PTOLEMY:2019hkd}, we define:
\begin{align}\label{eq:n_nu_loc}
  n_\nu^{\rm loc} = n_{\nu,0}(1+f_\mathrm{c}) \ ,
\end{align}
with $f_\mathrm{c}$ evaluated at the radius of the Earth inside the Milky Way. 

From the equations above we can clearly see that, as expected, the clustering is negligible for the High-$p_\nu$ case, while it is relevant for the Low-$T_\nu$+DR scenario. We should emphasise that our approach to obtain $f_\mathrm{c}$ for the Low-$T_\nu$+DR cosmology underestimates the actual clustering of neutrinos on Earth. For example, within $\Lambda$CDM, the linear method underestimates the clustering by a factor of ${\sim}3$ for $m_\nu \sim 1\,\text{eV}$, and by a factor of ${\sim}2$ for $m_\nu \sim 0.2\,\text{eV}$. In this regard, doing an N-body simulation for the non-standard cosmologies would be desirable, but lies beyond the scope of our work and will not affect the conclusions significantly. 
Indeed, within the context of PTOLEMY, our linear approach results in a smaller capture rate and therefore our predictions for these scenarios should be regarded as conservative.

To summarise this section, we have discussed how the stringent neutrino mass bound in $\Lambda$CDM can be evaded in two simple non-standard settings and estimated the number density of neutrinos expected on Earth for each of them. With this aspect of the calculation established, we can now turn to computing the relevant number of events and spectral features of the $\nu_e + {}^3{\rm H} \to e^- + {}^3{\rm He}^{+}$ processes in a PTOLEMY-like experiment.

\section{PTOLEMY Description and Phenomenology}
\label{sec:pheno}

\noindent In this section, we will outline our approach in obtaining the sensitivity of PTOLEMY to the Cosmic Neutrino Background. As usual with $\beta$-decay experiments, there are three separate contributions to the number of events that PTOLEMY will register: \emph{i)} the inverse $\beta$-capture of relic neutrinos, \emph{ii)} the electrons from the decay of the sample atoms, and \emph{iii)} additional background events from other sources. In this work, we adopt the same approach as in~\cite{PTOLEMY:2019hkd} to describe the contribution of the three processes to the total number of events. In what follows, we will highlight some of the calculations that are relevant for our purpose and refer the reader to~\cite{PTOLEMY:2019hkd} for the details.\\

\subsection{Event Rates}
\noindent The capture rate of relic neutrinos by tritium nuclei in Eq.~\eqref{eq:beta_capture} can be written as:
\begin{align}\label{eq:gamma_i}
    \Gamma_i = N_\mathrm{T}|U_{ei}|^2 \overline{\sigma}v_\nu \, n_{\nu}^{\rm loc}\ ,
\end{align}
with $i=\{1,2,3\}$ the index of the $i$th neutrino mass eigenstate, 
$|U_{ei}|^2$ the PMNS-matrix element,
$N_\mathrm{T}$ the number of tritium nuclei in the sample, and 
$\overline{\sigma}v_\nu \approx 3.8\times 10^{-45}\,{\rm cm}^2$ the interaction cross-section (see, e.g.,~\cite{Long:2014zva}). The local relic neutrino number density $n_{\nu}^{\rm loc}$ is related to the background neutrino number density $n_{\nu,0}$ via the gravitational clustering factor $f_\mathrm{c}$, as shown in Eq.~\eqref{eq:n_nu_loc}. 
In the mass region of interest for us here, we can always assume that neutrinos are quasi-degenerate and non-relativistic today. Under these conditions, all three mass states will cluster in the same way, such that $f_\mathrm{c}$ -- and thus $n_\nu^{\rm loc}$ -- is independent of the index $i$. 
Note that $n_\nu^{\rm loc}$ and $n_{\nu,0}$ are the number densities per helicity state. In the case of fully non-relativistic neutrinos, there will be twice as many left-chiral interacting states available for Majorana neutrinos than for Dirac neutrinos~\cite{Long:2014zva,Roulet:2018fyh,Akita:2020jbo}. Summing over the neutrino mass states, we have the following expression for the total relic neutrino event rate:
\begin{align} \label{eq:n_nu_loc-Gamma_CNB}
    \Gamma_{\rm CNB} &= c_{\rm D/M} \sum_i \Gamma_i  
    \approx  4 \,c_{\rm D/M} \,
    \frac{n_\nu^{\rm loc}}{n_{\nu,0}^{\rm SM}}
    \left(\frac{M_\mathrm{T}}{100\,{\rm g}}\right) 
     \,{\rm yr}^{-1} \ ,
\end{align}
where $c_{\rm D/M} = 1\,(2)$ for Dirac (Majorana) neutrinos, and $M_\mathrm{T} \equiv m_{{}^{3}\mathrm{H}}N_\mathrm{T}$ denotes the mass of the tritium sample (see below). This small event rate illustrates one of the main challenges of this type of measurement.

In order to estimate the sensitivity of the PTOLEMY experiment, we have to consider the electron energy spectrum.
Assuming non-relativistic relic neutrinos, kinematics imply that we should expect a monochromatic signal for each neutrino mass state:
\begin{align}
    \frac{\mathrm{d}\Gamma_i}{\mathrm{d}E_e} = \Gamma_i \delta\left[E_e-(E_\mathrm{end}^{m_\nu=0}+m_i)\right] \ ,
\end{align}
where $E_\mathrm{end}^{m_\nu=0}$ is the endpoint energy of the electron spectrum for massless neutrinos:
\begin{align}
    \label{eq:Eend_mnu0}
    E_\mathrm{end}^{m_\nu=0} = \frac{m_{{}^{3}\mathrm{H}}^2 + m_e^2 - m_{{}^{3}\mathrm{He}}^2}{2 m_{{}^{3}\mathrm{H}}}\ ,
\end{align}
with $m_{{}^{3}\mathrm{H}} \approx 2808.921\, \mathrm{MeV}$ and $m_{{}^{3}\mathrm{He}} \approx 2808.391\,\mathrm{MeV}$ the masses of the tritium and helium nuclei, respectively. Summing over the neutrino mass states and convolving with a Gaussian energy resolution of the detector leads to the total CNB induced rate:
\begin{align}
    \label{eq:Gamma_CNB}
     \frac{\mathrm{d}\widetilde{\Gamma}_\mathrm{CNB}}{\mathrm{d}E_e} &=  \frac{c_{\rm D/M}}{\sqrt{2\pi}(\Delta/\sqrt{8\ln(2)})}
    \nonumber\\
     &\times \sum_i  \Gamma_i 
     \exp\left\{-\frac{\left[E_e-(E_\mathrm{end}^{m_\nu=0}+m_i)\right]^2}{2\left(\Delta/\sqrt{8\ln(2)}\right)^2}\right\}\ ,
\end{align}
where $\Delta$ is the energy resolution.

The sample atoms can also decay into neutrinos and electrons, leading to a continuous $\beta$-decay background that completely dominates the total number of events below the endpoint electron energy $E_\mathrm{end} = E_\mathrm{end}^{m_\nu=0} - m_\mathrm{lightest}$. Nonetheless, given that the separation between the energies of background and signal electrons is roughly twice the neutrino mass, it could be possible to separate the two contributions if the energy resolution $\Delta$ is small enough compared to the neutrino mass. Analogously to Eq.~\eqref{eq:Gamma_CNB}, we can define a smoothed differential rate for the $\beta$-decay background as:
\begin{align}
    \label{eq:Gamma_background_smoothed}
    \frac{\mathrm{d} \widetilde{\Gamma}_{\beta}}{\mathrm{d} E_{e}}=&\frac{1}{\sqrt{2 \pi}(\Delta / \sqrt{8 \ln 2})} \int \mathrm{d} E^{\prime} \frac{\mathrm{d} \Gamma_{\beta}}{\mathrm{d} E^{\prime}} \times\nonumber\\
    & \times \exp \left[-\frac{\left(E_{e}-E^{\prime}\right)^{2}}{2(\Delta / \sqrt{8 \ln 2})^{2}}\right]\ ,
\end{align}
with
\begin{align}
    \label{eq:gamma_background}
    \frac{\mathrm{d} \Gamma_{\beta}}{\mathrm{d} E_{e}}=\frac{\overline{\sigma}}{\pi^{2}} N_\mathrm{T} \sum_{i=1}^{N_{\nu}}\left|U_{e i}\right|^{2} H\left(E_{e}, m_{i}\right)\ ,
\end{align}
where $H(E_{e}, m_{i})$ describes the continuous  $\beta$-spectrum and is given in Eq.~(3.9) of~\cite{PTOLEMY:2019hkd}. 

With these definitions, we can now write down the average number of expected background and signal events per energy bin $k$, centered at electron energy $E_k$ during some exposure time $T$:
\begin{align}
    \label{eq:N_events_background}
    N_{\beta}^{k} &=T \int_{E_{k}-\Delta / 2}^{E_{k}+\Delta / 2} \frac{d \widetilde{\Gamma}_{\beta}}{d E_{e}} d E_{e} \\
    \label{eq:N_events_CNB}
    N_{\mathrm{CNB}}^{k} &=T \int_{E_{k}-\Delta / 2}^{E_{k}+\Delta / 2} \frac{d \widetilde{\Gamma}_{\mathrm{CNB}}}{d E_{e}} d E_{e}\,,
\end{align}
where we set the size of the energy bins equal to the detector resolution $\Delta$. 
Following~\cite{PTOLEMY:2019hkd}, we consider electron energies in the range $-5\,\mathrm{eV} \leq E_e - E_\mathrm{end}^{m_\nu=0} \leq 10\,\mathrm{eV}$, such that the number of bins is given by $N_\mathrm{bins} = 15\,\mathrm{eV}/\Delta$. We have checked that our results for the PTOLEMY sensitivity change very little when reducing the analysis window to $-1.5\,\mathrm{eV} \leq E_e - E_\mathrm{end}^{m_\nu=0} \leq 1.5\,\mathrm{eV}$, for example.

In addition to $N^k_{\rm CNB}$ and $N^k_\beta$, we take into account background events from other sources, assuming that they contribute to an energy-independent rate $\Gamma_\mathrm{b}$. In our main analysis, we follow the approach in~\cite{PTOLEMY:2019hkd} and assume a background rate of $10^{-5}\, \mathrm{Hz}$ in the 15~eV analysis window, corresponding to $\Gamma_\mathrm{b} = 7\times 10^{-7} \,{\rm Hz/eV}$.
Moreover, we fix $T = 1\,\mathrm{yr}$, $\Delta = 100\,\mathrm{meV}$ and take a sample size of $M_\mathrm{T} = 100\,\mathrm{g}$. We investigate the dependence of our results on these experimental parameters in Sec.~\ref{sec:sensitivity}.

\subsection{Analysis Methodology}
\noindent We will now detail how we compute the sensitivity of PTOLEMY to the CNB. Firstly, we fix the values of exposure time $T$, sample mass $M_\mathrm{T}$, energy resolution $\Delta$, and local neutrino density $n_\nu^{\rm loc}$. In our main analysis, the values of the first three quantities are provided at the end of the previous subsection. Then, we define the total number of events per energy bin $k$ (our model prediction) as a function of the parameters $\boldsymbol{\theta}$ in our setup as:
\begin{align}
    \label{eq:N_model}
    &N^k(\boldsymbol{\theta}) = T \, \Delta \, \Gamma_\mathrm{b} \nonumber\\ 
    &+ A_\beta N_\beta^k(T,\Delta,M_\mathrm{T},m_\mathrm{lightest},\delta E_\mathrm{end}) \nonumber\\
    &+ A_\mathrm{CNB} N_\mathrm{CNB}^k(T,\Delta,M_\mathrm{T},n_\nu^\mathrm{loc}, m_\mathrm{lightest},\delta E_\mathrm{end})\ ,
\end{align}
where $\boldsymbol{\theta} = \{\Gamma_\mathrm{b}, A_\beta, m_\mathrm{lightest}, A_\mathrm{CNB}, \delta E_\mathrm{end}\}$ and in the first term we have used that the width of the energy bins is set to the resolution $\Delta$. Here, $A_\beta$ and $A_\mathrm{CNB}$ represent a normalisation factor in the number of background and signal events, respectively. The variable $\delta E_\mathrm{end}$ accounts for uncertainties in the endpoint energy of electrons $E_\mathrm{end}$. Next, to define our mock data, we choose fiducial values for these parameters:
\begin{align}
    \hat{\boldsymbol{\theta}} = \{\hat{\Gamma}_\mathrm{b}, \hat{A}_\beta = 1, \hat{m}_\mathrm{lightest}, \hat{A}_\mathrm{CNB} = 1, \widehat{\delta E}_\mathrm{end} = 0 \}\ ,
\end{align}
and construct a test statistic based on the Poisson likelihood:
\begin{align}
\label{eq:test_statistic}
\lambda(&\boldsymbol{\theta}; \hat{\boldsymbol{\theta}}) = 2\ln
\frac{\mathcal{L}(\hat{\boldsymbol{\theta}})}{\mathcal{L}(\boldsymbol{\theta})} \nonumber \\
&= 2\sum_k \left[ N^k(\boldsymbol{\theta}) - N^k(\hat{\boldsymbol{\theta}}) 
 + N^k(\hat{\boldsymbol{\theta}}) \ln \frac{N^k(\hat{\boldsymbol{\theta}})}{ N^k(\boldsymbol{\theta})}\right] \ ,
\end{align}
where we adopt the Asimov data set, given by the event numbers at the fiducial parameters $N^k(\hat{\boldsymbol{\theta}})$. Hence, the sensitivity obtained from this test statistic corresponds to the mean sensitivity over the possible statistical realisations. In order to test whether it is possible to establish the presence of the CNB within the chosen setup, we set $A_{\rm CNB} = 0$ in the parameter vector $\boldsymbol{\theta}$, and minimise $\lambda(\boldsymbol{\theta}; \hat{\boldsymbol{\theta}})$ with respect to all other parameters in $\boldsymbol{\theta}$. The sensitivity at 68.3\%, 95.5\%, 99.7\% confidence level (corresponding to 1, 2, 3 Gaussian standard deviations) is then obtained by  ${\min}\left[\lambda(\boldsymbol{\theta}; \hat{\boldsymbol{\theta}})\right] = 1,4,9$, respectively. We have explicitly checked that our approach is in excellent agreement with the Bayesian strategy, where a full MCMC analysis is employed to obtain the sensitivity of PTOLEMY to the CNB\footnote{The codes for both our frequentist and Bayesian approach can be found on the \href{https://github.com/james-alvey-42/DistNuAndPtolemy}{\texttt{GitHub}} page.}. Note that the sensitivity analysis presented by the PTOLEMY collaboration in Ref.~\cite{PTOLEMY:2019hkd} used the Bayesian method. We find good agreement with their results when adopting the same assumptions as in this reference. In the next section, we will use the described test statistic in Eq.~\eqref{eq:test_statistic} to calculate the sensitivity for CNB detection as a function of the local neutrino density $n_\nu^{\rm loc}$, lightest neutrino mass $m_{\rm lightest}$, and the experimental parameters $T$, $M_\mathrm{T}$, $\Delta$ and $\Gamma_{\rm b}$.

\begin{figure*}
    \centering
    \includegraphics[width=0.97\linewidth]{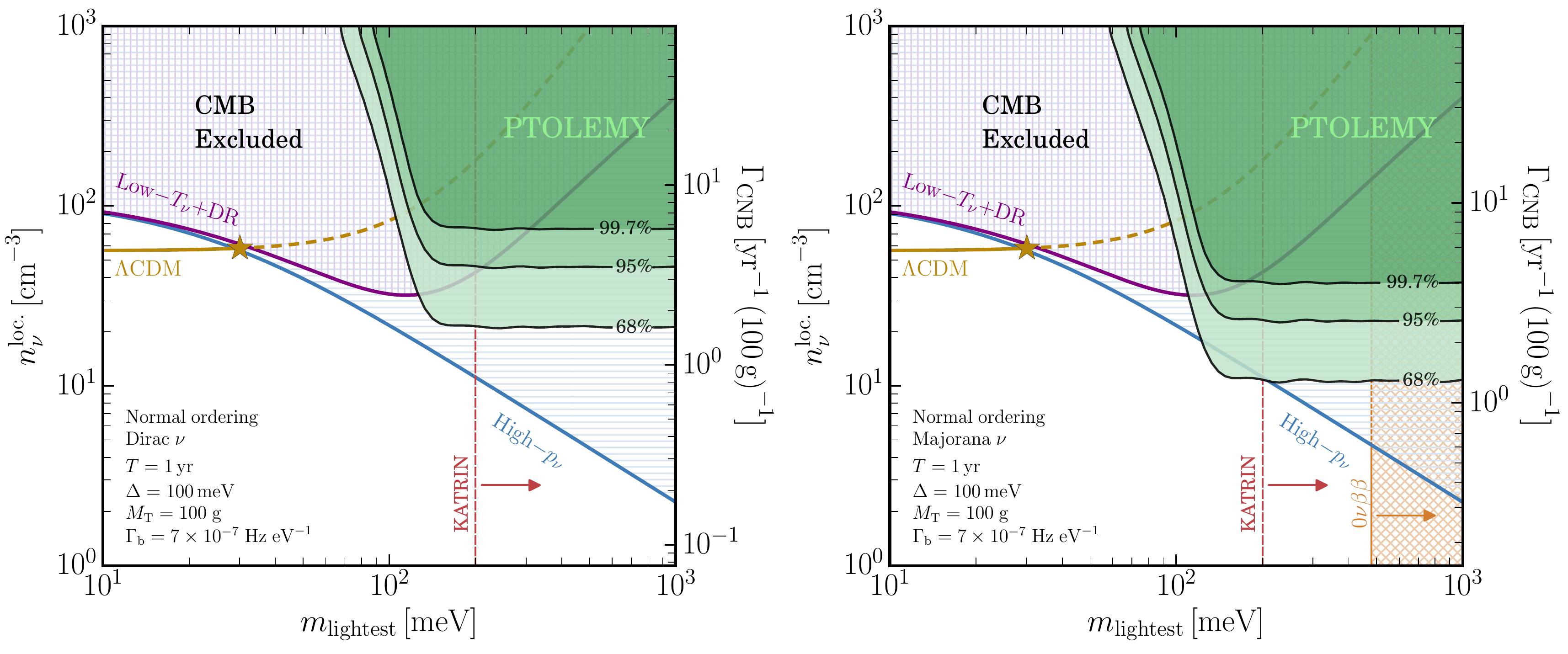}
    \caption{Sensitivity of a Cosmic Neutrino Background detection by PTOLEMY at 68\%, 95\%, and 99.7\%~CL for Dirac neutrinos (\emph{left}) and Majorana neutrinos (\emph{right}) as a function of the lightest neutrino mass $m_\mathrm{lightest}$ and the local neutrino density per helicity state $n_\nu^\mathrm{loc}$. The right vertical axis shows the CNB capture rate per year per $100\,\mathrm{g}$ of tritium sample mass. We assume a $1\,\mathrm{yr}$ exposure of $100\,\mathrm{g}$ tritium with an energy resolution of $100\,\mathrm{meV}$ and a background level of $7\times 10^{-7}\,\text{Hz}/\text{eV}$. The yellow curve shows the {\color{lcdmcol}{$\Lambda$CDM}} prediction including gravitational clustering in the Milky Way, and the star indicates the current upper bound on $m_{\rm lightest}$ from Planck, see Eq.~\eqref{eq:m0_planck}. The purple curve corresponds to the {\color{lowtcol}{Low-$T_\nu$+DR}} cosmology and the blue curve to the {\color{highpcol}{High-$p_\nu$}} scenario, see Section~\ref{sec:cosmopheno} for details. Here, the vertically-hatched and horizontally-hatched regions denote exclusion limits from CMB+BAO data in these cosmologies, respectively. We show the KATRIN sensitivity by a red arrow (see Eq.~\eqref{eq:Katrin_bound}). In the right panel we include the current limit from neutrinoless double-beta decay experiments, assuming the most conservative nuclear matrix element, see Eq.~\eqref{eq:mbb_bound}.}
    \label{fig:n_vs_mnu}
\end{figure*}

\section{Results}
\label{sec:results}

\subsection{PTOLEMY Sensitivity}
\label{sec:sensitivity}

\noindent In Fig.~\ref{fig:n_vs_mnu}, we show the parameter region where PTOLEMY could establish the presence of the CNB at 68\%, 95\%, and 99.7\%~CL if neutrinos are Dirac (left panel) or Majorana (right panel) particles. The sensitivity is shown as a function of the lightest neutrino mass and the local neutrino density per helicity state for our default experimental setup assumptions detailed above. The right axis show the corresponding relic neutrino capture rate per year, which is related to $n_\nu^{\rm loc}$ via Eq.~\eqref{eq:n_nu_loc-Gamma_CNB}. The factor 2 difference between Dirac and Majorana neutrinos leads to a corresponding shift between the left and right vertical axes in the two panels. Note that the values of $m_\mathrm{lightest}$ covered by the sensitivity region are large enough, such that the mass ordering (normal vs. inverted) is not so relevant here.

\begin{figure*}
    \centering
    \includegraphics[width=0.95\linewidth]{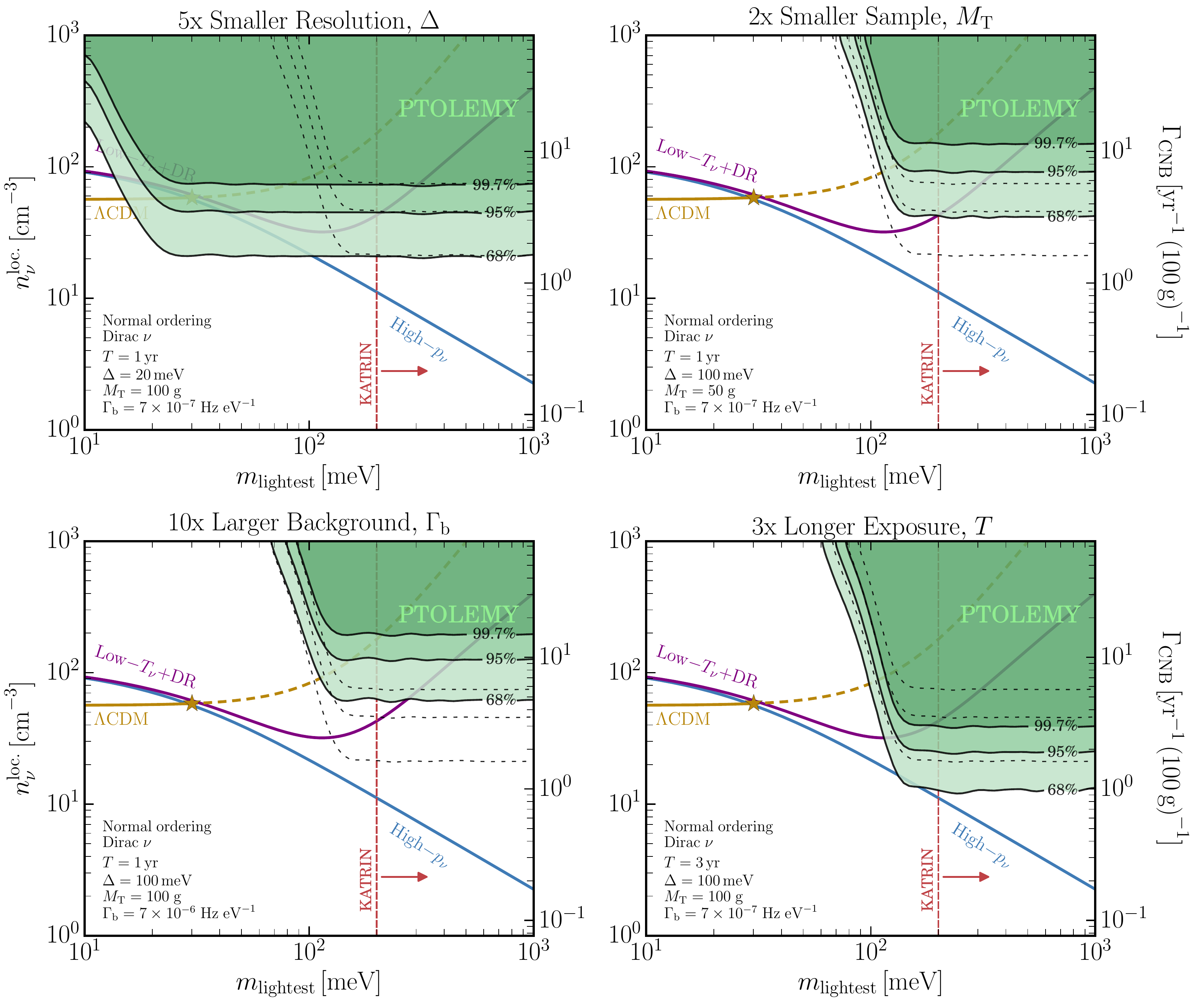}
    \caption{The dependence of the PTOLEMY sensitivity on the energy resolution (\emph{top left}), background rate (\emph{bottom left}), sample size (\emph{top right}), and exposure time (\emph{bottom right}). For comparison, the dotted curves correspond to the PTOLEMY sensitivity using our default parameters as in the left panel of Fig.~\ref{fig:n_vs_mnu}. Note that, as before, the regions above the purple and blue curves are excluded by CMB+BAO data.}\vspace{-12pt}
    \label{fig:PtolemyConfi}
\end{figure*}

The shape of the sensitivity region is easily understood: for small neutrino masses, the sensitivity diminishes because it becomes impossible to resolve the CNB peak from the $\beta$-decay continuum due to the finite-energy resolution of the detector, as illustrated in the left panel of Fig.~\ref{fig:spectrum}. Hence, the sensitivity limit towards small neutrino masses is controlled by the energy resolution $\Delta$: for values of $\Delta$ smaller than the $\Delta = 100$~meV assumed in Fig.~\ref{fig:n_vs_mnu}, the sensitivity region would extend to correspondingly smaller neutrino masses. In Fig.~\ref{fig:PtolemyConfi}, we show how the sensitivity of PTOLEMY depends on various experimental parameters. The behaviour with respect to the energy resolution mentioned here is illustrated in the top-left panel.

For low values of the neutrino number density, the induced signal becomes too small to be distinguished from the background events in the signal region. Therefore, the lower bound of the sensitivity region is set by the exposure time, sample mass and background rate, c.f., right panel of Fig.~\ref{fig:spectrum}. This effect is apparent by comparing the Dirac and Majorana cases in Fig.~\ref{fig:n_vs_mnu}, which differ by a factor two in $\Gamma_{\rm CNB}$ for fixed neutrino number density. This can also be seen by examining the top-right and bottom panels of Fig.~\ref{fig:PtolemyConfi}, where we highlight the dependence on $M_\mathrm{T}$, $\Gamma_\mathrm{b}$ and $T$.

To good accuracy, the behaviour of the flat, bottom part of the sensitivity region in the regime where $\Delta \ll m_\nu$ depends on the ratio signal/$\sqrt{{\rm background}}$. We find that the sensitivity at 68.3\%, 95.5\% and 99.7\%~CL in this regime can be written in terms of $n_\nu^{\rm loc}$ in a simple form as a function of $M_\mathrm{T}$, $\Gamma_\mathrm{b}$ and $T$:
\begin{align}
    \label{eq:analytic_sensitivity}
    \left[\frac{n_\nu^\mathrm{loc}}{56 \, \mathrm{cm}^{-3}}\right] &> \frac{\mathcal{A}}{c_\mathrm{D/M}} \left[\frac{100\,\mathrm{g}}{M_\mathrm{T}}\right] \left[\frac{1\,\mathrm{yr}}{T}\right]^\frac{1}{2}\left[\frac{\Gamma_\mathrm{b}}{7\times 10^{-7}\,\mathrm{Hz/eV}}\right]^\frac{1}{2} \nonumber \\ \nonumber\\
    \mathcal{A} &= \begin{cases} 0.380 &(68.3\%\ \mathrm{CL}) \\ 0.820 &(95.5\%\ \mathrm{CL}) \\ 1.332  &(99.7\%\ \mathrm{CL}) \\ \end{cases}\ .
\end{align}
The different $M_\mathrm{T}$ and $T$ dependencies follow by noting that both the CNB signal and the background events (as characterised by $\Gamma_\mathrm{b}$) scale linearly with $T$, see Eqs.~\eqref{eq:N_events_CNB} and~\eqref{eq:N_model}, whereas only the signal count depends on the sample mass $M_\mathrm{T}$. This equation provides a useful and efficient way to readily obtain a sensitivity estimate (i.e., how large $n_\nu^\mathrm{loc}$ should be to measure the CNB at the given CL) in the regime $\Delta \ll m_\nu$, without having to run a full analysis.

\subsection{Possibility of CNB Detection within the Experimental Landscape}

\begin{figure*}[t]
    \centering
    \includegraphics[width=\linewidth]{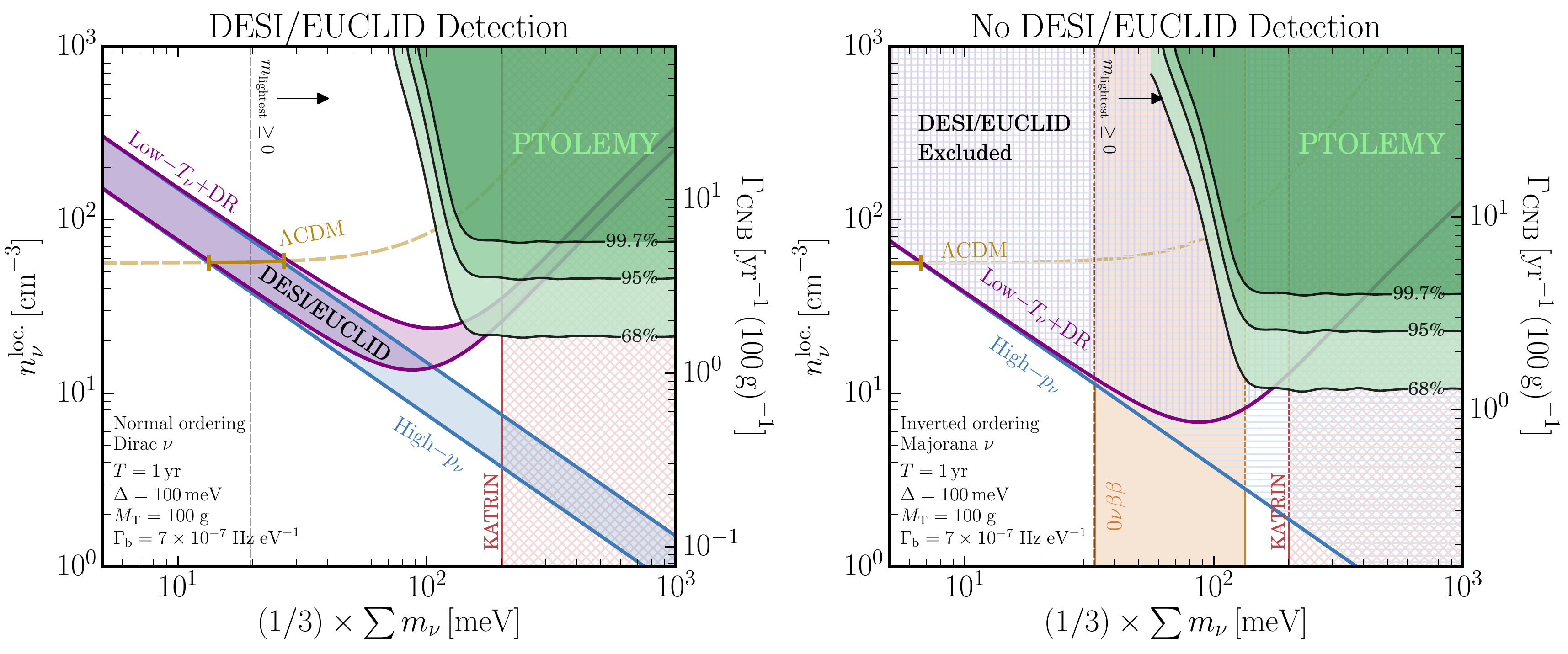}
    \caption{Possible future experimental scenarios. In the left panel, we assume Dirac neutrinos with normal-mass ordering and a cosmological neutrino mass determination of $(60\pm 20)\,{\rm meV}$ when interpreted within $\Lambda$CDM, see Eq.~\eqref{eq:numass-future}. In the right panel, we assume Majorana neutrinos with inverted ordering, an upper bound on the neutrino mass from cosmology corresponding to $\sum m_\nu < 20\,\mathrm{meV}$ in $\Lambda$CDM (see Eq.~\eqref{eq:numass-future2}), and a positive detection of $m_{\beta\beta} \approx 20-40\,\mathrm{meV}$ in neutrinoless double-beta decay experiments, corresponding to $100\,\text{meV} \lesssim  \sum m_\nu \lesssim 400\,\text{meV}$ (vertical orange-shaded band). Instead of showing $m_{\rm lightest}$ on the horizontal axes, we are using $\sum m_\nu/3$ here. The yellow line corresponds to the {\color{lcdmcol}{$\Lambda$CDM}} case, whilst the purple and blue lines represent the {\color{lowtcol}{Low-$T_\nu$+DR}} and {\color{highpcol}{High-$p_\nu$}} cosmologies, respectively. In the left panel, the shaded bands correspond to allowed regions, and in the right panel the hatched regions are excluded. Note that in this figure we have assumed that KATRIN has reached its final sensitivity (the hatched region to the right of the red, vertical line).}\vspace{-6pt}
    \label{fig:n_vs_mnu_future}
\end{figure*}

\noindent In addition to the PTOLEMY sensitivity, we show in Figs.~\ref{fig:n_vs_mnu} and \ref{fig:PtolemyConfi} the predictions of the three cosmological scenarios discussed in Sec.~\ref{sec:cosmopheno}. The yellow curve corresponds to the standard $\Lambda$CDM prediction. For light neutrino masses, the curve converges to $n_\nu^{\rm loc} \approx n_{\nu,0} \approx 56\,{\rm cm}^{-3}$, whereas for $m_{\rm lightest} \gtrsim 50\,\mathrm{meV}$ the number density increases due to the gravitational clustering in the Milky Way. The star corresponds to the Planck upper bound on $m_{\rm lightest}$ in Eq.~\eqref{eq:m0_planck}, and the dashed part of the yellow curve is excluded by cosmological observations within the $\Lambda$CDM scenario. We see that under the adopted assumptions, PTOLEMY will not be able to observe the CNB in this case, since for the assumed energy resolution, $m_{\rm lightest} \gtrsim 100$~meV is required. In order to reach the currently allowed masses within $\Lambda$CDM, the energy resolution has to be significantly higher and reach levels of $\Delta \sim 20\,{\rm meV}$. If such a good resolution can be achieved, then only a modest increase of the exposure time beyond $100\,\mathrm{g}\,\mathrm{yr}$ would be sufficient to reach a $3\sigma$ CNB detection in the case of Dirac neutrinos and under the background assumption of $7\times 10^{-7}$~Hz/eV, see top-left panel in Fig.~\ref{fig:PtolemyConfi}. It should be noted, however, that such a good resolution is likely extremely challenging to achieve from a technological viewpoint.

The purple and blue curves correspond to our two benchmark non-standard cosmologies, the Low-$T_\nu$+DR case and the High-$p_\nu$ scenario, respectively (see Sec.~\ref{sec:cosmopheno}). They both intersect the $\Lambda$CDM curve at the Planck bound on the sum of neutrino masses, as indicated by the yellow star. As such, these curves denote realisations of each model where the current cosmological bound on the non-relativistic neutrino energy density is saturated (Eq.~\eqref{eq:rhoNR_bound}), extending out to larger masses than those in $\Lambda$CDM. The regions above the purple and blue curves are excluded by current cosmological observations. Importantly, whilst the curves for these two scenarios coincide for small neutrino masses, they diverge for $m_{\rm lightest} \gtrsim 50$~meV where gravitational clustering becomes relevant, particularly in the Low-$T_\nu$+DR cosmology. This is in line with the discussion in Sec.~\ref{sec:cosmopheno}, where we argued that the clustering in the High-$p_\nu$ scenario is negligible. 

Focusing on the region $m_{\rm lightest} \gtrsim 200\,\mathrm{meV}$ -- corresponding to the projected sensitivity of KATRIN -- we see that it will still be very difficult to observe the CNB with PTOLEMY in the High-$p_\nu$ scenario, which is characterised by a strong suppression of the local relic neutrino density. However, for the Low-$T_\nu$+DR scenario, CNB detection prospects are more promising due to significant clustering: exposures of $200\, (100)\,\mathrm{g}\,\mathrm{yr}$ for Dirac (Majorana) neutrinos would enable a CNB detection at $>3\sigma$ for the energy resolution and background assumptions adopted in Fig.~\ref{fig:n_vs_mnu}. 

Let us now consider possible scenarios that could happen in the near future, when results from DESI/EUCLID and next-generation neutrinoless double-beta decay experiments will be available, and KATRIN will have reached its final sensitivity goal. We illustrate two representative cases in Fig.~\ref{fig:n_vs_mnu_future}. In the left panel, we assume that cosmological observations measure a finite value for the neutrino mass when interpreted within the $\Lambda$CDM framework. As discussed in Sec.~\ref{sec:cosmopheno}, such a measurement effectively determines the product $\sum m_\nu (2n_{\nu,0})$. For illustration purposes, we assume a cosmological neutrino energy density determination of: 
\begin{align}\label{eq:numass-future}
    \sum m_\nu \left(\frac{n_{\nu,0}}{56\,{\rm cm}^{-3}}\right) = (60 \pm 20) \,{\rm meV} \ ,
\end{align}
corresponding to the minimum allowed value for normal neutrino mass ordering in the $\Lambda$CDM model. This case is shown as the short, solid, yellow line in Fig.~\ref{fig:n_vs_mnu_future}. For the non-standard scenarios, the purple and blue bands will remain allowed, opening up the possibility for large values of the neutrino mass. If KATRIN by then had observed a finite neutrino mass within the hatched region in Fig.~\ref{fig:n_vs_mnu_future}, then the detection prospects of the Low-$T_\nu$+DR cosmology at PTOLEMY would be promising, while it would still be difficult to detect the CNB within the High-$p_\nu$ cosmology. If KATRIN would only set an upper limit, however, then CNB detection would be rather challenging in the two non-standard scenarios considered here, and require a combination of higher energy resolution, larger exposure, and lower background than assumed in Fig.~\ref{fig:n_vs_mnu_future}.

In the right panel of Fig.~\ref{fig:n_vs_mnu_future}, we assume Majorana neutrinos with an inverted-mass ordering and show a scenario where cosmology obtains only an upper bound on the neutrino energy density of:
\begin{align}\label{eq:numass-future2}
    \sum m_\nu \left(\frac{n_{\nu,0}}{56\,{\rm cm}^{-3}}\right)< 20\, {\rm meV} \ . 
\end{align}
Such a bound would be inconsistent with the minimal value for $\sum m_\nu$ required from oscillation data when interpreted within $\Lambda$CDM (see vertical, black line), since for the inverted-mass ordering, oscillation data bounds $\sum m_\nu \gtrsim 100\,\mathrm{meV}$. Therefore, such a result from cosmology would actually require a non-standard explanation. Furthermore, we indicate in the figure bounds from next-generation neutrinoless double-beta decay experiments, assuming that they will be consistent with oscillation data in the inverted-mass ordering~\cite{Giuliani:2019uno} and will determine $m_{\beta\beta} \approx 20-40$~meV, where we have assumed a factor of 2 uncertainty from nuclear matrix elements. Using Eq.~\eqref{eq:mbb} and marginalising over Majorana phases, this interval can be translated into $100\,{\rm meV} \lesssim \sum m_\nu \lesssim 400\,\mathrm{meV}$, as indicated by the vertical band in this panel. We observe that in such a scenario, KATRIN should obtain a null-result and a CNB detection will require substantial improvements (in energy resolution, exposure time and background rates) compared to our default assumptions.

In order to contextualise the implications of the futuristic scenarios shown in Fig.~\ref{fig:n_vs_mnu_future}, we believe that a timeline of expected experimental developments could be useful. Firstly, in the case of kinematic neutrino mass measurements, it is expected that KATRIN will reach its final sensitivity goal of $m_\beta \sim 0.2\,{\rm eV}$ in 3$-$4 years from now~\cite{KATRIN:2019yun,Aker:2021gma}. On a longer timescale, Project-8 is aiming to reach $m_\beta \sim 0.04\,{\rm eV}$ sensitivity by the end of this decade~\cite{Project8:2017nal}. Secondly, in terms of cosmological developments, the DESI experiment is already taking data and EUCLID should be launched soon. It is expected that the sensitivities shown in the left panel of Fig.~\ref{fig:n_vs_mnu_future} will be reached in ${\sim}4-5$ years from now. In addition, complementary cosmological probes, such as CMB observations by the Simons Observatory~\cite{SimonsObservatory:2018koc}, will improve the sensitivity to the energy density in non-relativistic neutrinos. Ultimately, this would simply reduce the size of the DESI/EUCLID contours shown in the left panel of Fig.~\ref{fig:n_vs_mnu_future}. Thirdly, the next-generation of neutrinoless double-beta decay experiments could be expected to start testing the inverted ordering scenario in $\sim 10\,{\rm years}$. It would likely take several more years to fully test the inverted-ordering case, once ton-scale experiments with very reduced backgrounds are operational~\cite{Giuliani:2019uno}. Finally, and importantly in the context of our study, one could hope that a PTOLEMY-like experiment could be developed and functional in 10$-$20 years.

\section{Landscape of Cosmological Scenarios}
\label{sec:scenarios}

\begin{table*}[t]
    \centering
\resizebox{\textwidth}{!}{
{\def\arraystretch{1.0}
    \begin{tabular}{l|l|p{8.2cm}}
    \hline\hline
\textbf{Cosmological Scenario} & \textbf{CNB Detection?} & \textbf{Comments}  \\ \hline \hline
\multirow{2}{*}{$\Lambda$CDM} & \multirow{2}{*}{Very challenging, see Figs.~\ref{fig:n_vs_mnu},~\ref{fig:PtolemyConfi}  and~\ref{fig:n_vs_mnu_future}}  &   \multirow{2}{=}{Similar considerations apply to cosmologies where neutrino properties are not drastically altered.} \\
   &   &   \\ \hline
Low-$T_\nu$+DR~\cite{Farzan:2015pca,GAMBITCosmologyWorkgroup:2020htv}  (see Sec.~\ref{sec:cosmopheno}) & Good, see magenta lines in Figs.~\ref{fig:n_vs_mnu}$-$\ref{fig:n_vs_mnu_future}  &  The detection prospects depend upon $T_\nu$ and $m_\nu$.  \\
High-$p_\nu$~\cite{Alvey:2021sji} (see Sec.~\ref{sec:cosmopheno}) & Unlikely, see blue lines in Figs.~\ref{fig:n_vs_mnu}$-$\ref{fig:n_vs_mnu_future} & There are very few neutrinos on Earth. \\ \hline \hline
Neutrino Decays, $\nu_i \to \nu_4 \,\phi$~\cite{Chacko:2019nej,Chacko:2020hmh,Escudero:2020ped,Abellan:2021rfq} &  No & No relic active neutrino background today. \\
Neutrino Decays, $\nu_i \to \nu_j \,\phi$~\cite{Escudero:2020ped,Barenboim:2020vrr,Akita:2021hqn} &  Yes, but only if $\Delta \lesssim 50\,\text{meV}$${}^\star$ & Only cosmologically viable for $\sum m_\nu \lesssim 0.17\,\text{eV}$~\cite{Escudero:2020ped}. \\ \hline 
Long range $\nu$ interactions~\cite{Esteban:2021ozz} & Potentially yes, with $\Delta \lesssim m_\nu$ & \multirow{2}{=}{The CNB would be made out of dense neutrino nuggets, but their distribution in the Universe is unknown~\cite{Ayaita:2014una,Casas:2016duf}.}    \\
Growing Neutrino Quintessence~\cite{Amendola:2007yx,Wetterich:2007kr,Pettorino:2010bv} & Potentially yes, with $\Delta \lesssim m_\nu$ &  \\ \hline
Neutrino masses from the $\theta$ term~\cite{Dvali:2016uhn,Dvali:2021uvk}& No & No relic neutrino background today.  \\
Neutrino masses from a late PT~\cite{Lorenz:2018fzb,Lorenz:2021alz}& Yes, provided that $\Delta \lesssim m_\nu$ & Only cosmologically viable for $\sum m_\nu < 1.41\,\text{eV}$~\cite{Lorenz:2021alz}. \\ 
\hline \hline
\end{tabular}
}
}
    \caption{Detection prospects of the CNB by a PTOLEMY-like experiment in a number of different cosmologies. Apart from $\Lambda$CDM, these cosmologies can accommodate for large neutrino masses, while still being in agreement with known cosmological data. Note that in our main analysis we have focused on the first three scenarios in the table. ${}^\star$In this case, both the neutrino mass $m_\nu$ and lifetime $\tau_\nu$ could potentially be inferred/constrained via a CNB detection, see Ref.~\cite{Akita:2021hqn}.}\vspace{-12pt}
    \label{tab:TheoryModels}
\end{table*}

\noindent In the previous sections, we have detailed the experimental requirements for a successful detection of the Cosmic Neutrino Background at a PTOLEMY-like experiment. We have seen that a key parameter controlling the detection prospects is the neutrino mass, which should be larger than the energy resolution of the detector, see e.g. Fig.~\ref{fig:spectrum}. In this context, besides the $\Lambda$CDM model, we have considered in detail the CNB detection prospects of two alternative cosmological models. As discussed in Sec.~\ref{sec:cosmopheno}, in these non-standard cosmologies the cosmological neutrino number density is smaller than in $\Lambda$CDM, which in turn means that the neutrino mass can be substantially higher and a CNB detection more achievable. However, as we have mentioned in the introduction, there exist several other non-standard cosmologies that can accommodate neutrinos with large masses while being in perfect agreement with all known cosmological data. In this section, we briefly discuss the reach of PTOLEMY for a number of alternative non-standard cosmologies, with the aim of understanding the cosmological implications of a CNB search at PTOLEMY. We summarise our main conclusions for these scenarios in Tab.~\ref{tab:TheoryModels}.

Essentially, non-standard cosmologies that can accommodate neutrinos with a large mass reduce, in one way or another, the energy density in non-relativistic neutrinos with respect to $\Lambda$CDM. The difference between each of these cosmologies is the way in which this is achieved\footnote{Note that typical extensions of the $\Lambda$CDM model that do not affect the neutrino sector, such as those featuring a dynamical dark energy equation of state or a non-standard primordial fluctuation spectrum, do not lead to a relevant reduction of the $\sum m_\nu$ bound with respect to $\Lambda$CDM, see e.g.~\cite{RoyChoudhury:2019hls,DiValentino:2019dzu}.}. For example, the cosmologies we considered in Sec.~\ref{sec:cosmopheno} involve neutrinos that have a distribution function that differs from the one within the Standard Model, which in turn allows them to have a smaller number density and therefore a smaller non-relativistic energy density. However, there are other ways to reduce the energy density of non-relativistic neutrinos in the Universe:

\begin{itemize}
    \item \textit{Decaying neutrinos}. Neutrinos could decay on cosmological timescales~\cite{Chacko:2019nej,Escudero:2020ped,Escudero:2019gfk,Barenboim:2020vrr,Chacko:2020hmh,Abellan:2021rfq}. In these scenarios, the cosmological neutrino mass bound can be relaxed because the decay product will be lighter than the neutrino itself. As a consequence, the extent to which the cosmological neutrino mass bound can be relaxed depends significantly on the final-state product. It has been shown that if all neutrinos decay into massless and inert species beyond the Standard Model ($\nu_i \to \nu_4 \,\phi$)~\cite{Escudero:2020ped}, then the cosmological neutrino mass bound can be relaxed up to $\sum m_\nu < 0.42\,\text{eV}$ at 95\% CL for lifetimes $\tau_\nu/t_\mathrm{Universe} \sim 10^{-4}-10^{-1}$~\cite{Abellan:2021rfq}. However, if this is the case, then there should be no relic neutrinos left today, which implies that no CNB detection will be possible in this scenario. Alternatively, neutrinos could decay into lighter active neutrino states by emitting a light boson $(\nu_i \to \nu_j \,\phi)$. In this case, there would still be a cosmic relic neutrino background. On the other hand, the cosmological neutrino mass bound would not be substantially relaxed in this case, because the final-state neutrinos have a mass that is only smaller by the mass splitting in Eqs.~\eqref{eq:mass_splittings_1} and~\eqref{eq:mass_splittings_2}. As such, while there will be a CNB to detect, the mass bound will be only moderately weakened, $\sum m_\nu \lesssim 0.17\,\text{eV}$~\cite{Escudero:2020ped}. In this mass range, the process of neutrino decay could in turn lead to interesting phenomenology in a PTOLEMY-like experiment, see e.g.~\cite{Akita:2021hqn} -- although it should be noted that detection would require a significantly more ambitious experimental configuration than the one assumed in the previous section. 

\item \textit{Long-range neutrino interactions}. Another way to reduce the energy density of non-relativistic neutrinos in the early Universe involves neutrinos that posses long-range interactions mediated by a very light scalar field~\cite{Esteban:2021ozz}, see also~\cite{Amendola:2007yx,Wetterich:2007kr,Pettorino:2010bv} and~\cite{Fardon:2003eh,Peccei:2004sz}. The main idea here is that, as a result of the long-range interactions, neutrinos behave as a massless fluid until temperatures $T \ll m_\nu$. In this way, the cosmological impact of their masses is substantially reduced. In particular, for scalar fields that do not behave as dark energy (with a mass $10^{-25} \lesssim m_\phi \lesssim 10^{-8}\,\text{eV}$), it has been shown that the neutrino mass could be as large as ${\sim}1\,\text{eV}$ if long-range interactions are present~\cite{Esteban:2021ozz}. On the other hand, if the scalar field behaves as dark energy~\cite{Amendola:2007yx,Wetterich:2007kr}, then it appears that CMB data restricts the neutrino mass to be at most $m_\nu \lesssim 0.5\,\text{eV}$~\cite{Pettorino:2010bv}. Regardless of the role of the light scalar field, once neutrinos eventually become non-relativistic (presumably at a rather low redshift, $z \lesssim 20$), the long-range neutrino interactions are such that neutrinos aggregate in dense nuggets of size $L \sim m_\phi^{-1}$~\cite{Afshordi:2005ym}. The study of the formation and evolution of these nuggets is rather complicated as a result of non-linear dynamics~\cite{Ayaita:2014una,Casas:2016duf} -- for a recent analytical study see~\cite{Smirnov:2022sfo}. This prevents one from making a definite prediction regarding the number density of neutrinos on Earth ($n_\nu^{\rm loc}$), which is clearly a key element in understanding the detection prospects of the CNB. Therefore, within the context of the CNB, it would be interesting to try to understand the neutrino nugget formation and evolution in depth, so that $n_\nu^{\rm loc}$ can be robustly predicted in these scenarios.

\item \textit{Late-time phase transitions}. Finally, neutrinos could obtain their masses at very low redshifts due to a late-time phase transition at $T \ll m_\nu$~\cite{Dvali:2016uhn,Dvali:2021uvk,Lorenz:2018fzb,Lorenz:2021alz}. This would reduce the energy density in non-relativistic neutrinos in the early Universe and thus allow for $m_\nu$ as large as $\sum m_\nu < 1.41\,\text{eV}$ at 95\% CL~\cite{Lorenz:2021alz}. In this case, the composition of the Cosmic Neutrino Background depends significantly upon the details of the late-time phase transition. For example, in the currently only known particle physics model capable of realising this effect~\cite{Dvali:2016uhn,Dvali:2021uvk}, one expects that there are no cosmic neutrinos today as they would have annihilated into massless and inert beyond-the-Standard-Model bosons. Nevertheless, by invoking additional ingredients such as neutrino chemical potentials, it may be possible to circumvent this issue, see~\cite{Lorenz:2018fzb,Lorenz:2021alz}. As a final comment on these scenarios, it is worth noting the neutrino mass can only be relaxed significantly provided that the phase transition occurs at a rather low redshift, $z\lesssim 0.5$~\cite{Lorenz:2021alz}. This fact could have important consequences for the determination of $n_\nu^{\rm loc}$, as massive neutrinos may not have had time to cluster in the Milky Way in these setups. 
\end{itemize}

\noindent To summarise, there currently exist a number of non-standard cosmologies where neutrinos could have a mass substantially larger than the one allowed in $\Lambda$CDM. All these scenarios are in full agreement with cosmological data, but require important modifications to the neutrino sector. In this regard, KATRIN is a key experiment, since if it reported a neutrino mass detection, the attention would naturally shift towards such alternative scenarios. Our discussion in this section and the previous one highlights that a suitably sensitive CNB search could help distinguish between them at the laboratory, see Tab.~\ref{tab:TheoryModels} for a summary. Importantly, we would like to emphasise that currently there are only a handful of cosmological models that could accommodate neutrinos with a mass large enough such that a PTOLEMY-like experiment could discover the CNB. In this context, it would be interesting to develop new, theoretically well-motivated cosmological settings that lead to a similar phenomenology.

\vspace*{-6pt}\section{Summary and Conclusions}
\label{sec:conclusions}

\noindent The direct detection of the Cosmic Neutrino Background (CNB) would represent an outstanding achievement in cosmology and particle physics. At present, the best chance to detect the CNB appears to be via neutrino capture on $\beta$-decaying nuclei in a PTOLEMY-like experiment~\cite{PTOLEMY:2018jst,PTOLEMY:2019hkd}. The prospects of such a process strongly depend upon several experimental factors and the cosmological model assumed. In particular, the neutrino mass ($m_\nu$) and number density on Earth ($n_\nu^{\rm loc}$) play a crucial role. The neutrino mass controls the minimum energy resolution $\Delta$ needed for a successful CNB detection, $\Delta \lesssim m_\nu$, while $n_\nu^{\rm loc}$ controls the CNB signal rate, which in turn determines the required minimum exposure time and maximum background rate. 

Within the $\Lambda$CDM model, current cosmological observations constrain the neutrino mass to be very small, $\sum m_\nu \lesssim 0.12\,\text{eV}$, which makes the detection prospects of the CNB very challenging. However, there are several non-standard cosmologies that allow for larger neutrino masses, while still being in agreement with all known cosmological data. As such, in this paper we have studied the CNB detection prospects in the context of cosmologies with large neutrino masses, as compared to the $\Lambda$CDM limit. For this purpose, in Sec.~\ref{sec:cosmopheno}, we have discussed in detail two example cosmologies featuring non-standard neutrino populations: one with neutrinos that have a smaller temperature than in the SM, $T_\nu < T_\nu^{\rm SM}$, supplemented with dark radiation, and one where neutrinos have a higher momentum than in $\Lambda$CDM, $\left<p_\nu \right> > 3.15\,T_\nu^{\rm SM}$. In each case, we have highlighted the cosmologically allowed ranges for $m_\nu$ and estimated the value of $n_\nu^{\rm loc}$ as a function of the neutrino mass. In Sec.~\ref{sec:pheno}, we have outlined our calculation of the CNB sensitivity at a PTOLEMY-like experiment. In Sec.~\ref{sec:results}, we have presented our results and discussed the experimental sensitivity that would be required to detect the CNB in the context of $\Lambda$CDM and the other non-standard cosmologies. Importantly, we have also contextualised the CNB detection prospects in light of KATRIN, next-generation neutrinoless double-beta decay experiments, and upcoming/future galaxy surveys by facilities such as DESI and EUCLID. Our main results are shown in Figs.~\ref{fig:n_vs_mnu},~\ref{fig:PtolemyConfi}, and~\ref{fig:n_vs_mnu_future}. Our main findings and conclusions can be summarised as follows:

\begin{itemize}
    \item \emph{CNB detection sensitivity:} We have computed the sensitivity of a PTOLEMY-like experiment within the general parameter space spanned by $(m_\mathrm{lightest}, n_\nu^\mathrm{loc})$. 
    Our numerical results can be qualitatively understood by a simple signal/$\sqrt{\text{background}}$ argument, which allows us to accurately estimate the sensitivity in terms of the required energy resolution, exposure, and background rate. Our results clearly highlight the fact that detection of the CNB within the context of $\Lambda$CDM is extremely challenging and would require an energy resolution $\Delta \lesssim 20 \, \mathrm{meV}$. On the other hand, in some of the large mass cosmologies considered here, the prospects are significantly improved for resolutions $\Delta \sim 100 \, \mathrm{meV}$.

    \item \emph{Experimental landscape:} There is an exciting current and future experimental program aimed at measuring the neutrino mass in the laboratory as well as in cosmology. 
    From a cosmological point of view, and within the standard $\Lambda$CDM model, upcoming galaxy surveys by DESI and EUCLID are expected to be sensitive to the minimal value of neutrino masses allowed by neutrino oscillation experiments. Concerning laboratory experiments, we have focused on the kinematic mass determination by KATRIN, as well as effective Majorana mass searches in neutrinoless double-beta decay facilities. Indeed, given the stringent upper limits from cosmology  within the $\Lambda$CDM scenario, it is unlikely that these experiments will be able to observe a positive signal. On the other hand, the situation in non-standard cosmologies is qualitatively different, allowing for large neutrino masses that are in reach of these laboratory experiments. Our analysis highlights that a CNB detection by a PTOLEMY-like experiment could be an important tool in distinguishing between different cosmological models, were there to be a positive detection in a laboratory neutrino mass experiment, or if cosmological limits would be seemingly in conflict with oscillation data when interpreted in the $\Lambda$CDM model.     
    
    \item \emph{Public code:} Alongside this paper, we have released a public code that can compute the CNB detection prospects of a PTOLEMY-like experiment within a cosmological scenario, given a specific experimental setup, neutrino mass $m_\mathrm{lightest}$ and local number density $n_\nu^\mathrm{loc}$. In addition, the code also provides a linear estimate of the gravitational clustering factor $f_\mathrm{c}$ for an arbitrary neutrino distribution function (see Eq.~\eqref{eq:n_nu_loc}), which is crucial in a number of relevant cosmological scenarios. The code is available at the following \href{https://github.com/james-alvey-42/DistNuAndPtolemy}{\texttt{GitHub}} page.
\end{itemize}

\noindent While the phenomenology of large neutrino masses paints an interesting picture at future facilities, an important part of the outlook is to understand the experimental challenges that face proposals such as PTOLEMY. For example, due to the significant technical challenge of isolating pure atomic tritium, the PTOLEMY collaboration suggested the idea to instead adhere tritium nuclei to graphene sheets~\cite{PTOLEMY:2018jst}. Given that the position of the tritium is then localised, this has important implications for the momentum-space behaviour of the final-state electron and helium as a result of the Heisenberg uncertainty principle~\cite{Cheipesh:2021fmg, Nussinov:2021zrj}. Even though there is currently no established analysis implementing this, it is likely that the result of this will be a decrease in the effective energy resolution. Ultimately, more research is required to understand this issue quantitatively, and it may lead to an update in the experimental setup, e.g., moving towards heavier nuclei~\cite{Mikulenko:2021ydo,Brdar:2022wuv}. 

As far as our analysis in this paper is concerned, however, we expect that somewhat irrespective of the setup, the key intuition will stay intact. In particular, provided that the effective energy resolution of the detector $\Delta$ is suitably small compared to the neutrino mass, the prospects for CNB detection will be similar to those derived in this work. Indeed, our semi-analytic estimate for the sensitivity in the large-mass regime will be broadly unchanged, updated only to account for modifications in, e.g., the capture cross-section. As such, we believe that regardless of the final design of a PTOLEMY-style detector, the exciting prospects to learn about neutrino properties with this proposed experiment will remain.

\vspace*{-12pt}
\section*{Acknowledgements}

\noindent JA is supported through the research program ``The Hidden Universe of Weakly Interacting Particles" with project number 680.92.18.03 (NWO Vrije Programma), which is partly financed by the Nederlandse Organisatie voor Wetenschappelijk Onderzoek (Dutch Research Council). ME is supported by a Fellowship of the Alexander von Humboldt
Foundation. NS is a recipient of a King's College London NMS Faculty Studentship. This project has received support from the European Union’s Horizon 2020 research and innovation program under the Marie Sklodowska-Curie grant agreement No 860881-HIDDeN.

\appendix
\vspace*{-6pt}\section{Neutrino Number Density on Earth} 
\label{app:nu_clustering}

\noindent The local number density of neutrinos is typically higher than the cosmological one, because of their gravitational clustering in the Milky-Way halo. In this appendix, we briefly describe our approach in obtaining the gravitational clustering factor $f_\mathrm{c}$ (which enters the rate in Eq.~\eqref{eq:gamma_i}, see also Eq.~\eqref{eq:n_nu_loc}) for a given cosmological distribution function of neutrinos. We closely follow the methods in~\cite{Ringwald:2004np}, where the clustering factor is estimated within linear theory. In principle, the full enhancement in the number density of neutrinos ought to be obtained through N-body simulations that capture the non-linear dynamics at small scales. Recently, there has been significant progress along these lines, where the effects of e.g. baryonic components and nearby clusters have been accounted for, see~\cite{Zhang:2017ljh,deSalas:2017wtt,Mertsch:2019qjv}. Nevertheless, the linear analysis gives a smaller estimate of $f_{\mathrm{c}}$ compared to the N-body result, which means that our derived PTOLEMY sensitivities are conservative in this regard. 

A full derivation of the clustering factor can be found in Sec.~6 of~\cite{Ringwald:2004np}. Here we only summarise the final result of the computation. The neutrino density contrast $\hat{\delta}_\nu(\mathrm{k}, s)$ is obtained as a solution of the collisionless Boltzmann equation and reads:
\begin{align}
    \label{eq:delta_nu_fourier}
    \hat{\delta}_{\nu}(\boldsymbol{k}, s) \simeq&\ 4 \pi G \overline{\rho}_\mathrm{m, 0} \int_{s_{i}}^{s} d s^{\prime} a\left(s^{\prime}\right) \hat{\delta}_\mathrm{m}\left(\boldsymbol{k}, s^{\prime}\right)\left(s-s^{\prime}\right)\nonumber\\
    & \times F\left[\frac{k\left(s-s^{\prime}\right)}{m_{\nu}}\right]\ ,
\end{align}
where $\overline{\rho}_\mathrm{m, 0}$ is the energy density of matter today, $a$ is the scale factor (which we obtain from the \texttt{CLASS} code~\cite{Lesgourgues:2011re,Blas:2011rf}), $\hat{\delta}_\mathrm{m}$ is the matter density contrast, $s = \int a^{-2}\mathrm{d}t$ is a time variable, and $F$ is given by:
\begin{align}
    F(q) \equiv \frac{1}{\bar{n}_{\nu, 0}} \int d^{3} p e^{-i \boldsymbol{p} \cdot \boldsymbol{q}} f_{0}(p)\ .
\end{align}
Here, $f_0$ is the distribution function of neutrinos, e.g., a Fermi-Dirac distribution within the $\Lambda$CDM model. The Fourier transform of Eq.~\eqref{eq:delta_nu_fourier} can then be taken to obtain the real-space neutrino overdensity today at the position of the Earth inside the Milky Way. In other words, we can obtain $f_\mathrm{c} \equiv \delta_\nu(r = 8 \, \mathrm{kpc})$ evaluated at redshift $z = 0$.

As far as the halo density contrast $\delta_\mathrm{m} = \rho_\mathrm{dm}/\bar{\rho}_\mathrm{m}$ in Eq.~\eqref{eq:delta_nu_fourier} is concerned, we follow the approach shared across Refs.~\cite{deSalas:2017wtt,Zhang:2017ljh,Mertsch:2019qjv}, which use updated data compared to the Milky-Way density profile given in appendix A.1 of~\cite{Ringwald:2004np}. Specifically, we take a general NFW-like profile for the Milky-Way dark-matter halo of the form:
\begin{equation}
    \rho_\mathrm{dm}(r, z) = \frac{\mathcal{N}(z)}{(r/r_\mathrm{s}(z))(1 + r/r_\mathrm{s}(z))^2}\ ,
\end{equation}
where we have noted explicitly the redshift dependence of the overall amplitude $\mathcal{N}$ and scale radius $r_\mathrm{s}$. The first step is then to take the $z = 0$ values given in table 1 of Ref.~\cite{Mertsch:2019qjv} for the scale radius $r_\mathrm{s}(0) = 19.9 \, \mathrm{kpc}$ and virial mass $M_\mathrm{vir} = 2.03 \times 10^{12} \, M_\odot$. Using the known expression for the density contrast at the virial radius for an NFW halo~\cite{Mertsch:2019qjv},
\begin{equation}
    \Delta_\mathrm{vir}(z) = 18\pi^2 + 82 \Omega_\mathrm{m}(z) - 39 \Omega_\mathrm{m}^2(z)\ ,
\end{equation}
we can compute the virial radius at $z = 0$ as:
\begin{equation}
    r_\mathrm{vir}(0) = \left(\frac{3M_\mathrm{vir}}{4\pi a_0^3 \Delta_\mathrm{vir}(0)\rho_{\mathrm{crit}, 0}}\right)^{1/3}\ .
\end{equation}
Here, $a_0 = 1$ and $\rho_{\mathrm{crit}, 0}$ are the scale factor and critical density today, respectively. This immediately gives the concentration parameter for the halo at $z = 0$, $c_\mathrm{vir}(0) = r_\mathrm{vir}(0)/r_\mathrm{s}(0)$, which is typically compared to an average value taken from simulations, $c_\mathrm{vir}(z) = \beta c_\mathrm{vir}^\mathrm{avg}(z)$, where $\beta$ is assumed to be a redshift-independent parameter and
\begin{equation}
    \log_{10} c_\mathrm{vir}^\mathrm{avg}(z) = a(z) + b(z) \log_{10}\left(\frac{M_\mathrm{vir}}{10^{12}\, h^{-1}M_\odot}\right)\ .
\end{equation}
In this expression, $a(z) = 0.537 + 0.488 \exp(-0.718 z^{1.08})$, $b(z) = -0.097 + 0.024 z$, and $h = 0.67$ is the Hubble parameter. In practice, this allows us to obtain the concentration parameter $c_\mathrm{vir}(z) = r_\mathrm{vir}(z)/r_\mathrm{s}(z)$ at any redshift after we compute $\beta$ using the $z = 0$ values for $c_\mathrm{vir}(0)$ and $c_\mathrm{vir}^\mathrm{avg}(0)$. This is now enough information to compute the evolution of $\mathcal{N}(z)$ and $r_\mathrm{s}(z)$. We can obtain the latter by first computing:
\begin{equation}
    r_\mathrm{vir}(z) = \left(\frac{3M_\mathrm{vir}}{4 \pi a^3 \Delta_\mathrm{vir}(z)\rho_\mathrm{crit}(z)}\right)^{1/3}\ ,
\end{equation}
and thus find $r_\mathrm{s}(z) = r_\mathrm{vir}(z)/c_\mathrm{vir}(z)$. Finally, we calculate the overall normalisation $\mathcal{N}(z)$ by choosing it to satisfy:
\begin{equation}
    M_\mathrm{vir} = 4\pi a^3 \int_0^{r_\mathrm{vir}(z)}{\mathrm{d}\tilde{r}\,\tilde{r}^2 \frac{\mathcal{N}(z)}{(r/r_\mathrm{s}(z))(1 + r/r_\mathrm{s}(z))^2}}\ .
\end{equation}
A full implementation of this scheme to compute the density profile and the neutrino density contrast (i.e., the gravitational clustering factor) for any neutrino distribution function can be found within the analysis section of the \href{https://github.com/james-alvey-42/DistNuAndPtolemy}{\texttt{GitHub}} page. 

In order to obtain the clustering factor in the cosmologies considered in Sec.~\ref{sec:cosmopheno}, we explicitly used the following distribution functions for neutrinos:

\begin{itemize}

\item  {\color{lcdmcol}{\boldmath{$\Lambda$}\textbf{CDM}}}: A Fermi-Dirac distribution with temperature $T_\nu^{\rm SM}$. 

\item {\color{lowtcol}{\textbf{Low-}\boldmath{$T_\nu$}\textbf{+DR}}}: A Fermi-Dirac distribution function with temperature $T_\nu = T_\nu^{\rm SM} \,(0.12\,{\rm eV}/ \sum m_\nu)^{1/3}$ that saturates the bound in Eq.~\eqref{eq:mnu_Tnudiff}. 

\item  {\color{highpcol}{\textbf{High-}\boldmath{$p_\nu$}}}: A Gaussian distribution function given by Eq.~(4) in~\cite{Alvey:2021sji}, with a mean $y_*$ and a width $\sigma_*$ that saturate the mass bound in Eq.~\eqref{eq:mnu_pggTnu}, and an amplitude $A$ such that $N_{\rm eff} = N_{\rm eff}^{\rm SM}$. We have explicitly checked that different combinations of $y_*$ and $\sigma_*$, satisfying the requirement above, all give $f_\mathrm{c} \simeq 0$, as claimed in the main text.
\end{itemize}

\bibliography{biblio}

\begin{thebibliography}{81}%
\makeatletter
\providecommand \@ifxundefined [1]{%
 \@ifx{#1\undefined}
}%
\providecommand \@ifnum [1]{%
 \ifnum #1\expandafter \@firstoftwo
 \else \expandafter \@secondoftwo
 \fi
}%
\providecommand \@ifx [1]{%
 \ifx #1\expandafter \@firstoftwo
 \else \expandafter \@secondoftwo
 \fi
}%
\providecommand \natexlab [1]{#1}%
\providecommand \enquote  [1]{``#1''}%
\providecommand \bibnamefont  [1]{#1}%
\providecommand \bibfnamefont [1]{#1}%
\providecommand \citenamefont [1]{#1}%
\providecommand \href@noop [0]{\@secondoftwo}%
\providecommand \href [0]{\begingroup \@sanitize@url \@href}%
\providecommand \@href[1]{\@@startlink{#1}\@@href}%
\providecommand \@@href[1]{\endgroup#1\@@endlink}%
\providecommand \@sanitize@url [0]{\catcode `\\12\catcode `\$12\catcode
  `\&12\catcode `\#12\catcode `\^12\catcode `\_12\catcode `\%12\relax}%
\providecommand \@@startlink[1]{}%
\providecommand \@@endlink[0]{}%
\providecommand \url  [0]{\begingroup\@sanitize@url \@url }%
\providecommand \@url [1]{\endgroup\@href {#1}{\urlprefix }}%
\providecommand \urlprefix  [0]{URL }%
\providecommand \Eprint [0]{\href }%
\providecommand \doibase [0]{http://dx.doi.org/}%
\providecommand \selectlanguage [0]{\@gobble}%
\providecommand \bibinfo  [0]{\@secondoftwo}%
\providecommand \bibfield  [0]{\@secondoftwo}%
\providecommand \translation [1]{[#1]}%
\providecommand \BibitemOpen [0]{}%
\providecommand \bibitemStop [0]{}%
\providecommand \bibitemNoStop [0]{.\EOS\space}%
\providecommand \EOS [0]{\spacefactor3000\relax}%
\providecommand \BibitemShut  [1]{\csname bibitem#1\endcsname}%
\let\auto@bib@innerbib\@empty
\bibitem [{\citenamefont {Dolgov}\ \emph {et~al.}(1997)\citenamefont {Dolgov},
  \citenamefont {Hansen},\ and\ \citenamefont {Semikoz}}]{Dolgov:1997mb}%
  \BibitemOpen
  \bibfield  {author} {\bibinfo {author} {\bibfnamefont {A.~D.}\ \bibnamefont
  {Dolgov}}, \bibinfo {author} {\bibfnamefont {S.~H.}\ \bibnamefont {Hansen}},
  \ and\ \bibinfo {author} {\bibfnamefont {D.~V.}\ \bibnamefont {Semikoz}},\
  }\href {\doibase 10.1016/S0550-3213(97)00479-3} {\bibfield  {journal}
  {\bibinfo  {journal} {Nucl. Phys. B}\ }\textbf {\bibinfo {volume} {503}},\
  \bibinfo {pages} {426} (\bibinfo {year} {1997})},\ \Eprint
  {http://arxiv.org/abs/hep-ph/9703315} {arXiv:hep-ph/9703315} \BibitemShut
  {NoStop}%
\bibitem [{\citenamefont {Mangano}\ \emph {et~al.}(2005)\citenamefont
  {Mangano}, \citenamefont {Miele}, \citenamefont {Pastor}, \citenamefont
  {Pinto}, \citenamefont {Pisanti},\ and\ \citenamefont
  {Serpico}}]{Mangano:2005cc}%
  \BibitemOpen
  \bibfield  {author} {\bibinfo {author} {\bibfnamefont {G.}~\bibnamefont
  {Mangano}}, \bibinfo {author} {\bibfnamefont {G.}~\bibnamefont {Miele}},
  \bibinfo {author} {\bibfnamefont {S.}~\bibnamefont {Pastor}}, \bibinfo
  {author} {\bibfnamefont {T.}~\bibnamefont {Pinto}}, \bibinfo {author}
  {\bibfnamefont {O.}~\bibnamefont {Pisanti}}, \ and\ \bibinfo {author}
  {\bibfnamefont {P.~D.}\ \bibnamefont {Serpico}},\ }\href {\doibase
  10.1016/j.nuclphysb.2005.09.041} {\bibfield  {journal} {\bibinfo  {journal}
  {Nucl. Phys. B}\ }\textbf {\bibinfo {volume} {729}},\ \bibinfo {pages} {221}
  (\bibinfo {year} {2005})},\ \Eprint {http://arxiv.org/abs/hep-ph/0506164}
  {arXiv:hep-ph/0506164} \BibitemShut {NoStop}%
\bibitem [{\citenamefont {de~Salas}\ and\ \citenamefont
  {Pastor}(2016)}]{deSalas:2016ztq}%
  \BibitemOpen
  \bibfield  {author} {\bibinfo {author} {\bibfnamefont {P.~F.}\ \bibnamefont
  {de~Salas}}\ and\ \bibinfo {author} {\bibfnamefont {S.}~\bibnamefont
  {Pastor}},\ }\href {\doibase 10.1088/1475-7516/2016/07/051} {\bibfield
  {journal} {\bibinfo  {journal} {JCAP}\ }\textbf {\bibinfo {volume} {07}},\
  \bibinfo {pages} {051} (\bibinfo {year} {2016})},\ \Eprint
  {http://arxiv.org/abs/1606.06986} {arXiv:1606.06986 [hep-ph]} \BibitemShut
  {NoStop}%
\bibitem [{\citenamefont {Escudero}(2019)}]{Escudero:2018mvt}%
  \BibitemOpen
  \bibfield  {author} {\bibinfo {author} {\bibfnamefont {M.}~\bibnamefont
  {Escudero}},\ }\href {\doibase 10.1088/1475-7516/2019/02/007} {\bibfield
  {journal} {\bibinfo  {journal} {JCAP}\ }\textbf {\bibinfo {volume} {02}},\
  \bibinfo {pages} {007} (\bibinfo {year} {2019})},\ \Eprint
  {http://arxiv.org/abs/1812.05605} {arXiv:1812.05605 [hep-ph]} \BibitemShut
  {NoStop}%
\bibitem [{\citenamefont {Pisanti}\ \emph {et~al.}(2021)\citenamefont
  {Pisanti}, \citenamefont {Mangano}, \citenamefont {Miele},\ and\
  \citenamefont {Mazzella}}]{Pisanti:2020efz}%
  \BibitemOpen
  \bibfield  {author} {\bibinfo {author} {\bibfnamefont {O.}~\bibnamefont
  {Pisanti}}, \bibinfo {author} {\bibfnamefont {G.}~\bibnamefont {Mangano}},
  \bibinfo {author} {\bibfnamefont {G.}~\bibnamefont {Miele}}, \ and\ \bibinfo
  {author} {\bibfnamefont {P.}~\bibnamefont {Mazzella}},\ }\href {\doibase
  10.1088/1475-7516/2021/04/020} {\bibfield  {journal} {\bibinfo  {journal}
  {JCAP}\ }\textbf {\bibinfo {volume} {04}},\ \bibinfo {pages} {020} (\bibinfo
  {year} {2021})},\ \Eprint {http://arxiv.org/abs/2011.11537} {arXiv:2011.11537
  [astro-ph.CO]} \BibitemShut {NoStop}%
\bibitem [{\citenamefont {Pitrou}\ \emph {et~al.}(2018)\citenamefont {Pitrou},
  \citenamefont {Coc}, \citenamefont {Uzan},\ and\ \citenamefont
  {Vangioni}}]{Pitrou:2018cgg}%
  \BibitemOpen
  \bibfield  {author} {\bibinfo {author} {\bibfnamefont {C.}~\bibnamefont
  {Pitrou}}, \bibinfo {author} {\bibfnamefont {A.}~\bibnamefont {Coc}},
  \bibinfo {author} {\bibfnamefont {J.-P.}\ \bibnamefont {Uzan}}, \ and\
  \bibinfo {author} {\bibfnamefont {E.}~\bibnamefont {Vangioni}},\ }\href
  {\doibase 10.1016/j.physrep.2018.04.005} {\bibfield  {journal} {\bibinfo
  {journal} {Phys. Rept.}\ }\textbf {\bibinfo {volume} {754}},\ \bibinfo
  {pages} {1} (\bibinfo {year} {2018})},\ \Eprint
  {http://arxiv.org/abs/1801.08023} {arXiv:1801.08023 [astro-ph.CO]}
  \BibitemShut {NoStop}%
\bibitem [{\citenamefont {Fields}\ \emph {et~al.}(2020)\citenamefont {Fields},
  \citenamefont {Olive}, \citenamefont {Yeh},\ and\ \citenamefont
  {Young}}]{Fields:2019pfx}%
  \BibitemOpen
  \bibfield  {author} {\bibinfo {author} {\bibfnamefont {B.~D.}\ \bibnamefont
  {Fields}}, \bibinfo {author} {\bibfnamefont {K.~A.}\ \bibnamefont {Olive}},
  \bibinfo {author} {\bibfnamefont {T.-H.}\ \bibnamefont {Yeh}}, \ and\
  \bibinfo {author} {\bibfnamefont {C.}~\bibnamefont {Young}},\ }\href
  {\doibase 10.1088/1475-7516/2020/03/010} {\bibfield  {journal} {\bibinfo
  {journal} {JCAP}\ }\textbf {\bibinfo {volume} {03}},\ \bibinfo {pages} {010}
  (\bibinfo {year} {2020})},\ \bibinfo {note} {[Erratum: JCAP 11, E02
  (2020)]},\ \Eprint {http://arxiv.org/abs/1912.01132} {arXiv:1912.01132
  [astro-ph.CO]} \BibitemShut {NoStop}%
\bibitem [{\citenamefont {Hsyu}\ \emph {et~al.}(2020)\citenamefont {Hsyu},
  \citenamefont {Cooke}, \citenamefont {Prochaska},\ and\ \citenamefont
  {Bolte}}]{Hsyu:2020uqb}%
  \BibitemOpen
  \bibfield  {author} {\bibinfo {author} {\bibfnamefont {T.}~\bibnamefont
  {Hsyu}}, \bibinfo {author} {\bibfnamefont {R.~J.}\ \bibnamefont {Cooke}},
  \bibinfo {author} {\bibfnamefont {J.~X.}\ \bibnamefont {Prochaska}}, \ and\
  \bibinfo {author} {\bibfnamefont {M.}~\bibnamefont {Bolte}},\ }\href
  {\doibase 10.3847/1538-4357/ab91af} {\bibfield  {journal} {\bibinfo
  {journal} {Astrophys. J.}\ }\textbf {\bibinfo {volume} {896}},\ \bibinfo
  {pages} {77} (\bibinfo {year} {2020})},\ \Eprint
  {http://arxiv.org/abs/2005.12290} {arXiv:2005.12290 [astro-ph.GA]}
  \BibitemShut {NoStop}%
\bibitem [{\citenamefont {Cooke}\ \emph {et~al.}(2018)\citenamefont {Cooke},
  \citenamefont {Pettini},\ and\ \citenamefont {Steidel}}]{Cooke:2017cwo}%
  \BibitemOpen
  \bibfield  {author} {\bibinfo {author} {\bibfnamefont {R.~J.}\ \bibnamefont
  {Cooke}}, \bibinfo {author} {\bibfnamefont {M.}~\bibnamefont {Pettini}}, \
  and\ \bibinfo {author} {\bibfnamefont {C.~C.}\ \bibnamefont {Steidel}},\
  }\href {\doibase 10.3847/1538-4357/aaab53} {\bibfield  {journal} {\bibinfo
  {journal} {Astrophys. J.}\ }\textbf {\bibinfo {volume} {855}},\ \bibinfo
  {pages} {102} (\bibinfo {year} {2018})},\ \Eprint
  {http://arxiv.org/abs/1710.11129} {arXiv:1710.11129 [astro-ph.CO]}
  \BibitemShut {NoStop}%
\bibitem [{\citenamefont {Mossa}\ \emph {et~al.}(2020)\citenamefont {Mossa}
  \emph {et~al.}}]{Mossa:2020gjc}%
  \BibitemOpen
  \bibfield  {author} {\bibinfo {author} {\bibfnamefont {V.}~\bibnamefont
  {Mossa}} \emph {et~al.},\ }\href {\doibase 10.1038/s41586-020-2878-4}
  {\bibfield  {journal} {\bibinfo  {journal} {Nature}\ }\textbf {\bibinfo
  {volume} {587}},\ \bibinfo {pages} {210} (\bibinfo {year}
  {2020})}\BibitemShut {NoStop}%
\bibitem [{\citenamefont {Aghanim}\ \emph {et~al.}(2020)\citenamefont {Aghanim}
  \emph {et~al.}}]{Aghanim:2018eyx}%
  \BibitemOpen
  \bibfield  {author} {\bibinfo {author} {\bibfnamefont {N.}~\bibnamefont
  {Aghanim}} \emph {et~al.} (\bibinfo {collaboration} {Planck}),\ }\href
  {\doibase 10.1051/0004-6361/201833910} {\bibfield  {journal} {\bibinfo
  {journal} {Astron. Astrophys.}\ }\textbf {\bibinfo {volume} {641}},\ \bibinfo
  {pages} {A6} (\bibinfo {year} {2020})},\ \Eprint
  {http://arxiv.org/abs/1807.06209} {arXiv:1807.06209 [astro-ph.CO]}
  \BibitemShut {NoStop}%
\bibitem [{\citenamefont {Escudero~Abenza}(2020)}]{EscuderoAbenza:2020cmq}%
  \BibitemOpen
  \bibfield  {author} {\bibinfo {author} {\bibfnamefont {M.}~\bibnamefont
  {Escudero~Abenza}},\ }\href {\doibase 10.1088/1475-7516/2020/05/048}
  {\bibfield  {journal} {\bibinfo  {journal} {JCAP}\ }\textbf {\bibinfo
  {volume} {05}},\ \bibinfo {pages} {048} (\bibinfo {year} {2020})},\ \Eprint
  {http://arxiv.org/abs/2001.04466} {arXiv:2001.04466 [hep-ph]} \BibitemShut
  {NoStop}%
\bibitem [{\citenamefont {Akita}\ and\ \citenamefont
  {Yamaguchi}(2020)}]{Akita:2020szl}%
  \BibitemOpen
  \bibfield  {author} {\bibinfo {author} {\bibfnamefont {K.}~\bibnamefont
  {Akita}}\ and\ \bibinfo {author} {\bibfnamefont {M.}~\bibnamefont
  {Yamaguchi}},\ }\href {\doibase 10.1088/1475-7516/2020/08/012} {\bibfield
  {journal} {\bibinfo  {journal} {JCAP}\ }\textbf {\bibinfo {volume} {08}},\
  \bibinfo {pages} {012} (\bibinfo {year} {2020})},\ \Eprint
  {http://arxiv.org/abs/2005.07047} {arXiv:2005.07047 [hep-ph]} \BibitemShut
  {NoStop}%
\bibitem [{\citenamefont {Bennett}\ \emph {et~al.}(2021)\citenamefont
  {Bennett}, \citenamefont {Buldgen}, \citenamefont {De~Salas}, \citenamefont
  {Drewes}, \citenamefont {Gariazzo}, \citenamefont {Pastor},\ and\
  \citenamefont {Wong}}]{Bennett:2020zkv}%
  \BibitemOpen
  \bibfield  {author} {\bibinfo {author} {\bibfnamefont {J.~J.}\ \bibnamefont
  {Bennett}}, \bibinfo {author} {\bibfnamefont {G.}~\bibnamefont {Buldgen}},
  \bibinfo {author} {\bibfnamefont {P.~F.}\ \bibnamefont {De~Salas}}, \bibinfo
  {author} {\bibfnamefont {M.}~\bibnamefont {Drewes}}, \bibinfo {author}
  {\bibfnamefont {S.}~\bibnamefont {Gariazzo}}, \bibinfo {author}
  {\bibfnamefont {S.}~\bibnamefont {Pastor}}, \ and\ \bibinfo {author}
  {\bibfnamefont {Y.~Y.~Y.}\ \bibnamefont {Wong}},\ }\href {\doibase
  10.1088/1475-7516/2021/04/073} {\bibfield  {journal} {\bibinfo  {journal}
  {JCAP}\ }\textbf {\bibinfo {volume} {04}},\ \bibinfo {pages} {073} (\bibinfo
  {year} {2021})},\ \Eprint {http://arxiv.org/abs/2012.02726} {arXiv:2012.02726
  [hep-ph]} \BibitemShut {NoStop}%
\bibitem [{\citenamefont {Froustey}\ \emph {et~al.}(2020)\citenamefont
  {Froustey}, \citenamefont {Pitrou},\ and\ \citenamefont
  {Volpe}}]{Froustey:2020mcq}%
  \BibitemOpen
  \bibfield  {author} {\bibinfo {author} {\bibfnamefont {J.}~\bibnamefont
  {Froustey}}, \bibinfo {author} {\bibfnamefont {C.}~\bibnamefont {Pitrou}}, \
  and\ \bibinfo {author} {\bibfnamefont {M.~C.}\ \bibnamefont {Volpe}},\ }\href
  {\doibase 10.1088/1475-7516/2020/12/015} {\bibfield  {journal} {\bibinfo
  {journal} {JCAP}\ }\textbf {\bibinfo {volume} {12}},\ \bibinfo {pages} {015}
  (\bibinfo {year} {2020})},\ \Eprint {http://arxiv.org/abs/2008.01074}
  {arXiv:2008.01074 [hep-ph]} \BibitemShut {NoStop}%
\bibitem [{\citenamefont {Weinberg}(1962)}]{Weinberg:1962zza}%
  \BibitemOpen
  \bibfield  {author} {\bibinfo {author} {\bibfnamefont {S.}~\bibnamefont
  {Weinberg}},\ }\href {\doibase 10.1103/PhysRev.128.1457} {\bibfield
  {journal} {\bibinfo  {journal} {Phys. Rev.}\ }\textbf {\bibinfo {volume}
  {128}},\ \bibinfo {pages} {1457} (\bibinfo {year} {1962})}\BibitemShut
  {NoStop}%
\bibitem [{\citenamefont {Cocco}\ \emph {et~al.}(2007)\citenamefont {Cocco},
  \citenamefont {Mangano},\ and\ \citenamefont {Messina}}]{Cocco:2007za}%
  \BibitemOpen
  \bibfield  {author} {\bibinfo {author} {\bibfnamefont {A.~G.}\ \bibnamefont
  {Cocco}}, \bibinfo {author} {\bibfnamefont {G.}~\bibnamefont {Mangano}}, \
  and\ \bibinfo {author} {\bibfnamefont {M.}~\bibnamefont {Messina}},\ }\href
  {\doibase 10.1088/1475-7516/2007/06/015} {\bibfield  {journal} {\bibinfo
  {journal} {JCAP}\ }\textbf {\bibinfo {volume} {06}},\ \bibinfo {pages} {015}
  (\bibinfo {year} {2007})},\ \Eprint {http://arxiv.org/abs/hep-ph/0703075}
  {arXiv:hep-ph/0703075} \BibitemShut {NoStop}%
\bibitem [{\citenamefont {Baracchini}\ \emph {et~al.}(2018)\citenamefont
  {Baracchini} \emph {et~al.}}]{PTOLEMY:2018jst}%
  \BibitemOpen
  \bibfield  {author} {\bibinfo {author} {\bibfnamefont {E.}~\bibnamefont
  {Baracchini}} \emph {et~al.} (\bibinfo {collaboration} {PTOLEMY}),\
  }\href@noop {} {\  (\bibinfo {year} {2018})},\ \Eprint
  {http://arxiv.org/abs/1808.01892} {arXiv:1808.01892 [physics.ins-det]}
  \BibitemShut {NoStop}%
\bibitem [{\citenamefont {Cheipesh}\ \emph {et~al.}(2021)\citenamefont
  {Cheipesh}, \citenamefont {Cheianov},\ and\ \citenamefont
  {Boyarsky}}]{Cheipesh:2021fmg}%
  \BibitemOpen
  \bibfield  {author} {\bibinfo {author} {\bibfnamefont {Y.}~\bibnamefont
  {Cheipesh}}, \bibinfo {author} {\bibfnamefont {V.}~\bibnamefont {Cheianov}},
  \ and\ \bibinfo {author} {\bibfnamefont {A.}~\bibnamefont {Boyarsky}},\
  }\href {\doibase 10.1103/PhysRevD.104.116004} {\bibfield  {journal} {\bibinfo
   {journal} {Phys. Rev. D}\ }\textbf {\bibinfo {volume} {104}},\ \bibinfo
  {pages} {116004} (\bibinfo {year} {2021})},\ \Eprint
  {http://arxiv.org/abs/2101.10069} {arXiv:2101.10069 [hep-ph]} \BibitemShut
  {NoStop}%
\bibitem [{\citenamefont {Nussinov}\ and\ \citenamefont
  {Nussinov}(2021)}]{Nussinov:2021zrj}%
  \BibitemOpen
  \bibfield  {author} {\bibinfo {author} {\bibfnamefont {S.}~\bibnamefont
  {Nussinov}}\ and\ \bibinfo {author} {\bibfnamefont {Z.}~\bibnamefont
  {Nussinov}},\ }\href@noop {} {\  (\bibinfo {year} {2021})},\ \Eprint
  {http://arxiv.org/abs/2108.03695} {arXiv:2108.03695 [hep-ph]} \BibitemShut
  {NoStop}%
\bibitem [{\citenamefont {Blennow}(2008)}]{Blennow:2008fh}%
  \BibitemOpen
  \bibfield  {author} {\bibinfo {author} {\bibfnamefont {M.}~\bibnamefont
  {Blennow}},\ }\href {\doibase 10.1103/PhysRevD.77.113014} {\bibfield
  {journal} {\bibinfo  {journal} {Phys. Rev. D}\ }\textbf {\bibinfo {volume}
  {77}},\ \bibinfo {pages} {113014} (\bibinfo {year} {2008})},\ \Eprint
  {http://arxiv.org/abs/0803.3762} {arXiv:0803.3762 [astro-ph]} \BibitemShut
  {NoStop}%
\bibitem [{\citenamefont {Long}\ \emph {et~al.}(2014)\citenamefont {Long},
  \citenamefont {Lunardini},\ and\ \citenamefont {Sabancilar}}]{Long:2014zva}%
  \BibitemOpen
  \bibfield  {author} {\bibinfo {author} {\bibfnamefont {A.~J.}\ \bibnamefont
  {Long}}, \bibinfo {author} {\bibfnamefont {C.}~\bibnamefont {Lunardini}}, \
  and\ \bibinfo {author} {\bibfnamefont {E.}~\bibnamefont {Sabancilar}},\
  }\href {\doibase 10.1088/1475-7516/2014/08/038} {\bibfield  {journal}
  {\bibinfo  {journal} {JCAP}\ }\textbf {\bibinfo {volume} {08}},\ \bibinfo
  {pages} {038} (\bibinfo {year} {2014})},\ \Eprint
  {http://arxiv.org/abs/1405.7654} {arXiv:1405.7654 [hep-ph]} \BibitemShut
  {NoStop}%
\bibitem [{\citenamefont {Betti}\ \emph {et~al.}(2019)\citenamefont {Betti}
  \emph {et~al.}}]{PTOLEMY:2019hkd}%
  \BibitemOpen
  \bibfield  {author} {\bibinfo {author} {\bibfnamefont {M.~G.}\ \bibnamefont
  {Betti}} \emph {et~al.} (\bibinfo {collaboration} {PTOLEMY}),\ }\href
  {\doibase 10.1088/1475-7516/2019/07/047} {\bibfield  {journal} {\bibinfo
  {journal} {JCAP}\ }\textbf {\bibinfo {volume} {07}},\ \bibinfo {pages} {047}
  (\bibinfo {year} {2019})},\ \Eprint {http://arxiv.org/abs/1902.05508}
  {arXiv:1902.05508 [astro-ph.CO]} \BibitemShut {NoStop}%
\bibitem [{\citenamefont {Akita}\ \emph
  {et~al.}(2021{\natexlab{a}})\citenamefont {Akita}, \citenamefont {Hurwitz},\
  and\ \citenamefont {Yamaguchi}}]{Akita:2020jbo}%
  \BibitemOpen
  \bibfield  {author} {\bibinfo {author} {\bibfnamefont {K.}~\bibnamefont
  {Akita}}, \bibinfo {author} {\bibfnamefont {S.}~\bibnamefont {Hurwitz}}, \
  and\ \bibinfo {author} {\bibfnamefont {M.}~\bibnamefont {Yamaguchi}},\ }\href
  {\doibase 10.1140/epjc/s10052-021-09133-5} {\bibfield  {journal} {\bibinfo
  {journal} {Eur. Phys. J. C}\ }\textbf {\bibinfo {volume} {81}},\ \bibinfo
  {pages} {344} (\bibinfo {year} {2021}{\natexlab{a}})},\ \Eprint
  {http://arxiv.org/abs/2010.04454} {arXiv:2010.04454 [hep-ph]} \BibitemShut
  {NoStop}%
\bibitem [{\citenamefont {Formaggio}\ \emph {et~al.}(2021)\citenamefont
  {Formaggio}, \citenamefont {de~Gouv\^ea},\ and\ \citenamefont
  {Robertson}}]{Formaggio:2021nfz}%
  \BibitemOpen
  \bibfield  {author} {\bibinfo {author} {\bibfnamefont {J.~A.}\ \bibnamefont
  {Formaggio}}, \bibinfo {author} {\bibfnamefont {A.~L.~C.}\ \bibnamefont
  {de~Gouv\^ea}}, \ and\ \bibinfo {author} {\bibfnamefont {R.~G.~H.}\
  \bibnamefont {Robertson}},\ }\href {\doibase 10.1016/j.physrep.2021.02.002}
  {\bibfield  {journal} {\bibinfo  {journal} {Phys. Rept.}\ }\textbf {\bibinfo
  {volume} {914}},\ \bibinfo {pages} {1} (\bibinfo {year} {2021})},\ \Eprint
  {http://arxiv.org/abs/2102.00594} {arXiv:2102.00594 [nucl-ex]} \BibitemShut
  {NoStop}%
\bibitem [{\citenamefont {Esteban}\ \emph {et~al.}(2020)\citenamefont
  {Esteban}, \citenamefont {Gonzalez-Garcia}, \citenamefont {Maltoni},
  \citenamefont {Schwetz},\ and\ \citenamefont {Zhou}}]{Esteban:2020cvm}%
  \BibitemOpen
  \bibfield  {author} {\bibinfo {author} {\bibfnamefont {I.}~\bibnamefont
  {Esteban}}, \bibinfo {author} {\bibfnamefont {M.}~\bibnamefont
  {Gonzalez-Garcia}}, \bibinfo {author} {\bibfnamefont {M.}~\bibnamefont
  {Maltoni}}, \bibinfo {author} {\bibfnamefont {T.}~\bibnamefont {Schwetz}}, \
  and\ \bibinfo {author} {\bibfnamefont {A.}~\bibnamefont {Zhou}},\ }\href
  {\doibase 10.1007/JHEP09(2020)178} {\bibfield  {journal} {\bibinfo  {journal}
  {JHEP}\ }\textbf {\bibinfo {volume} {09}},\ \bibinfo {pages} {178} (\bibinfo
  {year} {2020})},\ \Eprint {http://arxiv.org/abs/2007.14792} {arXiv:2007.14792
  [hep-ph]} \BibitemShut {NoStop}%
\bibitem [{\citenamefont {de~Salas}\ \emph {et~al.}(2021)\citenamefont
  {de~Salas}, \citenamefont {Forero}, \citenamefont {Gariazzo}, \citenamefont
  {Mart\'\i{}nez-Mirav\'e}, \citenamefont {Mena}, \citenamefont {Ternes},
  \citenamefont {T\'ortola},\ and\ \citenamefont {Valle}}]{deSalas:2020pgw}%
  \BibitemOpen
  \bibfield  {author} {\bibinfo {author} {\bibfnamefont {P.~F.}\ \bibnamefont
  {de~Salas}}, \bibinfo {author} {\bibfnamefont {D.~V.}\ \bibnamefont
  {Forero}}, \bibinfo {author} {\bibfnamefont {S.}~\bibnamefont {Gariazzo}},
  \bibinfo {author} {\bibfnamefont {P.}~\bibnamefont {Mart\'\i{}nez-Mirav\'e}},
  \bibinfo {author} {\bibfnamefont {O.}~\bibnamefont {Mena}}, \bibinfo {author}
  {\bibfnamefont {C.~A.}\ \bibnamefont {Ternes}}, \bibinfo {author}
  {\bibfnamefont {M.}~\bibnamefont {T\'ortola}}, \ and\ \bibinfo {author}
  {\bibfnamefont {J.~W.~F.}\ \bibnamefont {Valle}},\ }\href {\doibase
  10.1007/JHEP02(2021)071} {\bibfield  {journal} {\bibinfo  {journal} {JHEP}\
  }\textbf {\bibinfo {volume} {02}},\ \bibinfo {pages} {071} (\bibinfo {year}
  {2021})},\ \Eprint {http://arxiv.org/abs/2006.11237} {arXiv:2006.11237
  [hep-ph]} \BibitemShut {NoStop}%
\bibitem [{\citenamefont {Capozzi}\ \emph {et~al.}(2021)\citenamefont
  {Capozzi}, \citenamefont {Di~Valentino}, \citenamefont {Lisi}, \citenamefont
  {Marrone}, \citenamefont {Melchiorri},\ and\ \citenamefont
  {Palazzo}}]{Capozzi:2021fjo}%
  \BibitemOpen
  \bibfield  {author} {\bibinfo {author} {\bibfnamefont {F.}~\bibnamefont
  {Capozzi}}, \bibinfo {author} {\bibfnamefont {E.}~\bibnamefont
  {Di~Valentino}}, \bibinfo {author} {\bibfnamefont {E.}~\bibnamefont {Lisi}},
  \bibinfo {author} {\bibfnamefont {A.}~\bibnamefont {Marrone}}, \bibinfo
  {author} {\bibfnamefont {A.}~\bibnamefont {Melchiorri}}, \ and\ \bibinfo
  {author} {\bibfnamefont {A.}~\bibnamefont {Palazzo}},\ }\href@noop {} {\
  (\bibinfo {year} {2021})},\ \Eprint {http://arxiv.org/abs/2107.00532}
  {arXiv:2107.00532 [hep-ph]} \BibitemShut {NoStop}%
\bibitem [{\citenamefont {Aker}\ \emph {et~al.}(2019)\citenamefont {Aker} \emph
  {et~al.}}]{KATRIN:2019yun}%
  \BibitemOpen
  \bibfield  {author} {\bibinfo {author} {\bibfnamefont {M.}~\bibnamefont
  {Aker}} \emph {et~al.} (\bibinfo {collaboration} {KATRIN}),\ }\href {\doibase
  10.1103/PhysRevLett.123.221802} {\bibfield  {journal} {\bibinfo  {journal}
  {Phys. Rev. Lett.}\ }\textbf {\bibinfo {volume} {123}},\ \bibinfo {pages}
  {221802} (\bibinfo {year} {2019})},\ \Eprint
  {http://arxiv.org/abs/1909.06048} {arXiv:1909.06048 [hep-ex]} \BibitemShut
  {NoStop}%
\bibitem [{\citenamefont {Aker}\ \emph {et~al.}(2021)\citenamefont {Aker} \emph
  {et~al.}}]{Aker:2021gma}%
  \BibitemOpen
  \bibfield  {author} {\bibinfo {author} {\bibfnamefont {M.}~\bibnamefont
  {Aker}} \emph {et~al.} (\bibinfo {collaboration} {KATRIN}),\ }\href@noop {}
  {\  (\bibinfo {year} {2021})},\ \Eprint {http://arxiv.org/abs/2105.08533}
  {arXiv:2105.08533 [hep-ex]} \BibitemShut {NoStop}%
\bibitem [{\citenamefont {Ashtari~Esfahani}\ \emph {et~al.}(2017)\citenamefont
  {Ashtari~Esfahani} \emph {et~al.}}]{Project8:2017nal}%
  \BibitemOpen
  \bibfield  {author} {\bibinfo {author} {\bibfnamefont {A.}~\bibnamefont
  {Ashtari~Esfahani}} \emph {et~al.} (\bibinfo {collaboration} {Project 8}),\
  }\href {\doibase 10.1088/1361-6471/aa5b4f} {\bibfield  {journal} {\bibinfo
  {journal} {J. Phys. G}\ }\textbf {\bibinfo {volume} {44}},\ \bibinfo {pages}
  {054004} (\bibinfo {year} {2017})},\ \Eprint
  {http://arxiv.org/abs/1703.02037} {arXiv:1703.02037 [physics.ins-det]}
  \BibitemShut {NoStop}%
\bibitem [{\citenamefont {Gastaldo}\ \emph {et~al.}(2017)\citenamefont
  {Gastaldo} \emph {et~al.}}]{Gastaldo:2017edk}%
  \BibitemOpen
  \bibfield  {author} {\bibinfo {author} {\bibfnamefont {L.}~\bibnamefont
  {Gastaldo}} \emph {et~al.},\ }\href {\doibase 10.1140/epjst/e2017-70071-y}
  {\bibfield  {journal} {\bibinfo  {journal} {Eur. Phys. J. ST}\ }\textbf
  {\bibinfo {volume} {226}},\ \bibinfo {pages} {1623} (\bibinfo {year}
  {2017})}\BibitemShut {NoStop}%
\bibitem [{\citenamefont {Di~Valentino}\ \emph {et~al.}(2021)\citenamefont
  {Di~Valentino}, \citenamefont {Gariazzo},\ and\ \citenamefont
  {Mena}}]{DiValentino:2021hoh}%
  \BibitemOpen
  \bibfield  {author} {\bibinfo {author} {\bibfnamefont {E.}~\bibnamefont
  {Di~Valentino}}, \bibinfo {author} {\bibfnamefont {S.}~\bibnamefont
  {Gariazzo}}, \ and\ \bibinfo {author} {\bibfnamefont {O.}~\bibnamefont
  {Mena}},\ }\href {\doibase 10.1103/PhysRevD.104.083504} {\bibfield  {journal}
  {\bibinfo  {journal} {Phys. Rev. D}\ }\textbf {\bibinfo {volume} {104}},\
  \bibinfo {pages} {083504} (\bibinfo {year} {2021})},\ \Eprint
  {http://arxiv.org/abs/2106.15267} {arXiv:2106.15267 [astro-ph.CO]}
  \BibitemShut {NoStop}%
\bibitem [{\citenamefont {Aghamousa}\ \emph {et~al.}(2016)\citenamefont
  {Aghamousa} \emph {et~al.}}]{DESI:2016fyo}%
  \BibitemOpen
  \bibfield  {author} {\bibinfo {author} {\bibfnamefont {A.}~\bibnamefont
  {Aghamousa}} \emph {et~al.} (\bibinfo {collaboration} {DESI}),\ }\href@noop
  {} {\  (\bibinfo {year} {2016})},\ \Eprint {http://arxiv.org/abs/1611.00036}
  {arXiv:1611.00036 [astro-ph.IM]} \BibitemShut {NoStop}%
\bibitem [{\citenamefont {Amendola}\ \emph {et~al.}(2018)\citenamefont
  {Amendola} \emph {et~al.}}]{Amendola:2016saw}%
  \BibitemOpen
  \bibfield  {author} {\bibinfo {author} {\bibfnamefont {L.}~\bibnamefont
  {Amendola}} \emph {et~al.},\ }\href {\doibase 10.1007/s41114-017-0010-3}
  {\bibfield  {journal} {\bibinfo  {journal} {Living Rev. Rel.}\ }\textbf
  {\bibinfo {volume} {21}},\ \bibinfo {pages} {2} (\bibinfo {year} {2018})},\
  \Eprint {http://arxiv.org/abs/1606.00180} {arXiv:1606.00180 [astro-ph.CO]}
  \BibitemShut {NoStop}%
\bibitem [{\citenamefont {Brinckmann}\ \emph {et~al.}(2019)\citenamefont
  {Brinckmann}, \citenamefont {Hooper}, \citenamefont {Archidiacono},
  \citenamefont {Lesgourgues},\ and\ \citenamefont
  {Sprenger}}]{Brinckmann:2018owf}%
  \BibitemOpen
  \bibfield  {author} {\bibinfo {author} {\bibfnamefont {T.}~\bibnamefont
  {Brinckmann}}, \bibinfo {author} {\bibfnamefont {D.~C.}\ \bibnamefont
  {Hooper}}, \bibinfo {author} {\bibfnamefont {M.}~\bibnamefont
  {Archidiacono}}, \bibinfo {author} {\bibfnamefont {J.}~\bibnamefont
  {Lesgourgues}}, \ and\ \bibinfo {author} {\bibfnamefont {T.}~\bibnamefont
  {Sprenger}},\ }\href {\doibase 10.1088/1475-7516/2019/01/059} {\bibfield
  {journal} {\bibinfo  {journal} {JCAP}\ }\textbf {\bibinfo {volume} {01}},\
  \bibinfo {pages} {059} (\bibinfo {year} {2019})},\ \Eprint
  {http://arxiv.org/abs/1808.05955} {arXiv:1808.05955 [astro-ph.CO]}
  \BibitemShut {NoStop}%
\bibitem [{\citenamefont {Gando}\ \emph {et~al.}(2016)\citenamefont {Gando}
  \emph {et~al.}}]{KamLAND-Zen:2016pfg}%
  \BibitemOpen
  \bibfield  {author} {\bibinfo {author} {\bibfnamefont {A.}~\bibnamefont
  {Gando}} \emph {et~al.} (\bibinfo {collaboration} {KamLAND-Zen}),\ }\href
  {\doibase 10.1103/PhysRevLett.117.109903, 10.1103/PhysRevLett.117.082503}
  {\bibfield  {journal} {\bibinfo  {journal} {Phys. Rev. Lett.}\ }\textbf
  {\bibinfo {volume} {117}},\ \bibinfo {pages} {082503} (\bibinfo {year}
  {2016})},\ \bibinfo {note} {[Addendum: Phys. Rev.
  Lett.117,no.10,109903(2016)]},\ \Eprint {http://arxiv.org/abs/1605.02889}
  {arXiv:1605.02889 [hep-ex]} \BibitemShut {NoStop}%
\bibitem [{\citenamefont {Agostini}\ \emph {et~al.}(2020)\citenamefont
  {Agostini} \emph {et~al.}}]{GERDA:2020xhi}%
  \BibitemOpen
  \bibfield  {author} {\bibinfo {author} {\bibfnamefont {M.}~\bibnamefont
  {Agostini}} \emph {et~al.} (\bibinfo {collaboration} {GERDA}),\ }\href
  {\doibase 10.1103/PhysRevLett.125.252502} {\bibfield  {journal} {\bibinfo
  {journal} {Phys. Rev. Lett.}\ }\textbf {\bibinfo {volume} {125}},\ \bibinfo
  {pages} {252502} (\bibinfo {year} {2020})},\ \Eprint
  {http://arxiv.org/abs/2009.06079} {arXiv:2009.06079 [nucl-ex]} \BibitemShut
  {NoStop}%
\bibitem [{\citenamefont {Adams}\ \emph {et~al.}(2021)\citenamefont {Adams}
  \emph {et~al.}}]{CUORE:2021gpk}%
  \BibitemOpen
  \bibfield  {author} {\bibinfo {author} {\bibfnamefont {D.~Q.}\ \bibnamefont
  {Adams}} \emph {et~al.} (\bibinfo {collaboration} {CUORE}),\ }\href@noop {}
  {\  (\bibinfo {year} {2021})},\ \Eprint {http://arxiv.org/abs/2104.06906}
  {arXiv:2104.06906 [nucl-ex]} \BibitemShut {NoStop}%
\bibitem [{\citenamefont {Anton}\ \emph {et~al.}(2019)\citenamefont {Anton}
  \emph {et~al.}}]{EXO-200:2019rkq}%
  \BibitemOpen
  \bibfield  {author} {\bibinfo {author} {\bibfnamefont {G.}~\bibnamefont
  {Anton}} \emph {et~al.} (\bibinfo {collaboration} {EXO-200}),\ }\href
  {\doibase 10.1103/PhysRevLett.123.161802} {\bibfield  {journal} {\bibinfo
  {journal} {Phys. Rev. Lett.}\ }\textbf {\bibinfo {volume} {123}},\ \bibinfo
  {pages} {161802} (\bibinfo {year} {2019})},\ \Eprint
  {http://arxiv.org/abs/1906.02723} {arXiv:1906.02723 [hep-ex]} \BibitemShut
  {NoStop}%
\bibitem [{\citenamefont {Giuliani}\ \emph {et~al.}(2019)\citenamefont
  {Giuliani}, \citenamefont {Gomez~Cadenas}, \citenamefont {Pascoli},
  \citenamefont {Previtali}, \citenamefont {Saakyan}, \citenamefont
  {Sch\"affner},\ and\ \citenamefont {Sch\"onert}}]{Giuliani:2019uno}%
  \BibitemOpen
  \bibfield  {author} {\bibinfo {author} {\bibfnamefont {A.}~\bibnamefont
  {Giuliani}}, \bibinfo {author} {\bibfnamefont {J.~J.}\ \bibnamefont
  {Gomez~Cadenas}}, \bibinfo {author} {\bibfnamefont {S.}~\bibnamefont
  {Pascoli}}, \bibinfo {author} {\bibfnamefont {E.}~\bibnamefont {Previtali}},
  \bibinfo {author} {\bibfnamefont {R.}~\bibnamefont {Saakyan}}, \bibinfo
  {author} {\bibfnamefont {K.}~\bibnamefont {Sch\"affner}}, \ and\ \bibinfo
  {author} {\bibfnamefont {S.}~\bibnamefont {Sch\"onert}} (\bibinfo
  {collaboration} {APPEC Committee}),\ }\href@noop {} {\  (\bibinfo {year}
  {2019})},\ \Eprint {http://arxiv.org/abs/1910.04688} {arXiv:1910.04688
  [hep-ex]} \BibitemShut {NoStop}%
\bibitem [{\citenamefont {Alvey}\ \emph {et~al.}(2022)\citenamefont {Alvey},
  \citenamefont {Escudero},\ and\ \citenamefont {Sabti}}]{Alvey:2021sji}%
  \BibitemOpen
  \bibfield  {author} {\bibinfo {author} {\bibfnamefont {J.}~\bibnamefont
  {Alvey}}, \bibinfo {author} {\bibfnamefont {M.}~\bibnamefont {Escudero}}, \
  and\ \bibinfo {author} {\bibfnamefont {N.}~\bibnamefont {Sabti}},\ }\href
  {\doibase 10.1088/1475-7516/2022/02/037} {\bibfield  {journal} {\bibinfo
  {journal} {JCAP}\ }\textbf {\bibinfo {volume} {02}},\ \bibinfo {pages} {037}
  (\bibinfo {year} {2022})},\ \Eprint {http://arxiv.org/abs/2111.12726}
  {arXiv:2111.12726 [astro-ph.CO]} \BibitemShut {NoStop}%
\bibitem [{\citenamefont {Chacko}\ \emph {et~al.}(2020)\citenamefont {Chacko},
  \citenamefont {Dev}, \citenamefont {Du}, \citenamefont {Poulin},\ and\
  \citenamefont {Tsai}}]{Chacko:2019nej}%
  \BibitemOpen
  \bibfield  {author} {\bibinfo {author} {\bibfnamefont {Z.}~\bibnamefont
  {Chacko}}, \bibinfo {author} {\bibfnamefont {A.}~\bibnamefont {Dev}},
  \bibinfo {author} {\bibfnamefont {P.}~\bibnamefont {Du}}, \bibinfo {author}
  {\bibfnamefont {V.}~\bibnamefont {Poulin}}, \ and\ \bibinfo {author}
  {\bibfnamefont {Y.}~\bibnamefont {Tsai}},\ }\href {\doibase
  10.1007/JHEP04(2020)020} {\bibfield  {journal} {\bibinfo  {journal} {JHEP}\
  }\textbf {\bibinfo {volume} {04}},\ \bibinfo {pages} {020} (\bibinfo {year}
  {2020})},\ \Eprint {http://arxiv.org/abs/1909.05275} {arXiv:1909.05275
  [hep-ph]} \BibitemShut {NoStop}%
\bibitem [{\citenamefont {Escudero}\ \emph {et~al.}(2020)\citenamefont
  {Escudero}, \citenamefont {Lopez-Pavon}, \citenamefont {Rius},\ and\
  \citenamefont {Sandner}}]{Escudero:2020ped}%
  \BibitemOpen
  \bibfield  {author} {\bibinfo {author} {\bibfnamefont {M.}~\bibnamefont
  {Escudero}}, \bibinfo {author} {\bibfnamefont {J.}~\bibnamefont
  {Lopez-Pavon}}, \bibinfo {author} {\bibfnamefont {N.}~\bibnamefont {Rius}}, \
  and\ \bibinfo {author} {\bibfnamefont {S.}~\bibnamefont {Sandner}},\ }\href
  {\doibase 10.1007/JHEP12(2020)119} {\bibfield  {journal} {\bibinfo  {journal}
  {JHEP}\ }\textbf {\bibinfo {volume} {12}},\ \bibinfo {pages} {119} (\bibinfo
  {year} {2020})},\ \Eprint {http://arxiv.org/abs/2007.04994} {arXiv:2007.04994
  [hep-ph]} \BibitemShut {NoStop}%
\bibitem [{\citenamefont {Escudero}\ and\ \citenamefont
  {Fairbairn}(2019)}]{Escudero:2019gfk}%
  \BibitemOpen
  \bibfield  {author} {\bibinfo {author} {\bibfnamefont {M.}~\bibnamefont
  {Escudero}}\ and\ \bibinfo {author} {\bibfnamefont {M.}~\bibnamefont
  {Fairbairn}},\ }\href {\doibase 10.1103/PhysRevD.100.103531} {\bibfield
  {journal} {\bibinfo  {journal} {Phys. Rev. D}\ }\textbf {\bibinfo {volume}
  {100}},\ \bibinfo {pages} {103531} (\bibinfo {year} {2019})},\ \Eprint
  {http://arxiv.org/abs/1907.05425} {arXiv:1907.05425 [hep-ph]} \BibitemShut
  {NoStop}%
\bibitem [{\citenamefont {Barenboim}\ \emph {et~al.}(2021)\citenamefont
  {Barenboim}, \citenamefont {Chen}, \citenamefont {Hannestad}, \citenamefont
  {Oldengott}, \citenamefont {Tram},\ and\ \citenamefont
  {Wong}}]{Barenboim:2020vrr}%
  \BibitemOpen
  \bibfield  {author} {\bibinfo {author} {\bibfnamefont {G.}~\bibnamefont
  {Barenboim}}, \bibinfo {author} {\bibfnamefont {J.~Z.}\ \bibnamefont {Chen}},
  \bibinfo {author} {\bibfnamefont {S.}~\bibnamefont {Hannestad}}, \bibinfo
  {author} {\bibfnamefont {I.~M.}\ \bibnamefont {Oldengott}}, \bibinfo {author}
  {\bibfnamefont {T.}~\bibnamefont {Tram}}, \ and\ \bibinfo {author}
  {\bibfnamefont {Y.~Y.~Y.}\ \bibnamefont {Wong}},\ }\href {\doibase
  10.1088/1475-7516/2021/03/087} {\bibfield  {journal} {\bibinfo  {journal}
  {JCAP}\ }\textbf {\bibinfo {volume} {03}},\ \bibinfo {pages} {087} (\bibinfo
  {year} {2021})},\ \Eprint {http://arxiv.org/abs/2011.01502} {arXiv:2011.01502
  [astro-ph.CO]} \BibitemShut {NoStop}%
\bibitem [{\citenamefont {Chacko}\ \emph {et~al.}(2021)\citenamefont {Chacko},
  \citenamefont {Dev}, \citenamefont {Du}, \citenamefont {Poulin},\ and\
  \citenamefont {Tsai}}]{Chacko:2020hmh}%
  \BibitemOpen
  \bibfield  {author} {\bibinfo {author} {\bibfnamefont {Z.}~\bibnamefont
  {Chacko}}, \bibinfo {author} {\bibfnamefont {A.}~\bibnamefont {Dev}},
  \bibinfo {author} {\bibfnamefont {P.}~\bibnamefont {Du}}, \bibinfo {author}
  {\bibfnamefont {V.}~\bibnamefont {Poulin}}, \ and\ \bibinfo {author}
  {\bibfnamefont {Y.}~\bibnamefont {Tsai}},\ }\href {\doibase
  10.1103/PhysRevD.103.043519} {\bibfield  {journal} {\bibinfo  {journal}
  {Phys. Rev. D}\ }\textbf {\bibinfo {volume} {103}},\ \bibinfo {pages}
  {043519} (\bibinfo {year} {2021})},\ \Eprint
  {http://arxiv.org/abs/2002.08401} {arXiv:2002.08401 [astro-ph.CO]}
  \BibitemShut {NoStop}%
\bibitem [{\citenamefont {Abell\'an}\ \emph {et~al.}(2021)\citenamefont
  {Abell\'an}, \citenamefont {Chacko}, \citenamefont {Dev}, \citenamefont {Du},
  \citenamefont {Poulin},\ and\ \citenamefont {Tsai}}]{Abellan:2021rfq}%
  \BibitemOpen
  \bibfield  {author} {\bibinfo {author} {\bibfnamefont {G.~F.}\ \bibnamefont
  {Abell\'an}}, \bibinfo {author} {\bibfnamefont {Z.}~\bibnamefont {Chacko}},
  \bibinfo {author} {\bibfnamefont {A.}~\bibnamefont {Dev}}, \bibinfo {author}
  {\bibfnamefont {P.}~\bibnamefont {Du}}, \bibinfo {author} {\bibfnamefont
  {V.}~\bibnamefont {Poulin}}, \ and\ \bibinfo {author} {\bibfnamefont
  {Y.}~\bibnamefont {Tsai}},\ }\href@noop {} {\  (\bibinfo {year} {2021})},\
  \Eprint {http://arxiv.org/abs/2112.13862} {arXiv:2112.13862 [hep-ph]}
  \BibitemShut {NoStop}%
\bibitem [{\citenamefont {Dvali}\ and\ \citenamefont
  {Funcke}(2016)}]{Dvali:2016uhn}%
  \BibitemOpen
  \bibfield  {author} {\bibinfo {author} {\bibfnamefont {G.}~\bibnamefont
  {Dvali}}\ and\ \bibinfo {author} {\bibfnamefont {L.}~\bibnamefont {Funcke}},\
  }\href {\doibase 10.1103/PhysRevD.93.113002} {\bibfield  {journal} {\bibinfo
  {journal} {Phys. Rev. D}\ }\textbf {\bibinfo {volume} {93}},\ \bibinfo
  {pages} {113002} (\bibinfo {year} {2016})},\ \Eprint
  {http://arxiv.org/abs/1602.03191} {arXiv:1602.03191 [hep-ph]} \BibitemShut
  {NoStop}%
\bibitem [{\citenamefont {Dvali}\ \emph {et~al.}(2021)\citenamefont {Dvali},
  \citenamefont {Funcke},\ and\ \citenamefont {Vachaspati}}]{Dvali:2021uvk}%
  \BibitemOpen
  \bibfield  {author} {\bibinfo {author} {\bibfnamefont {G.}~\bibnamefont
  {Dvali}}, \bibinfo {author} {\bibfnamefont {L.}~\bibnamefont {Funcke}}, \
  and\ \bibinfo {author} {\bibfnamefont {T.}~\bibnamefont {Vachaspati}},\
  }\href@noop {} {\  (\bibinfo {year} {2021})},\ \Eprint
  {http://arxiv.org/abs/2112.02107} {arXiv:2112.02107 [hep-ph]} \BibitemShut
  {NoStop}%
\bibitem [{\citenamefont {Lorenz}\ \emph {et~al.}(2019)\citenamefont {Lorenz},
  \citenamefont {Funcke}, \citenamefont {Calabrese},\ and\ \citenamefont
  {Hannestad}}]{Lorenz:2018fzb}%
  \BibitemOpen
  \bibfield  {author} {\bibinfo {author} {\bibfnamefont {C.~S.}\ \bibnamefont
  {Lorenz}}, \bibinfo {author} {\bibfnamefont {L.}~\bibnamefont {Funcke}},
  \bibinfo {author} {\bibfnamefont {E.}~\bibnamefont {Calabrese}}, \ and\
  \bibinfo {author} {\bibfnamefont {S.}~\bibnamefont {Hannestad}},\ }\href
  {\doibase 10.1103/PhysRevD.99.023501} {\bibfield  {journal} {\bibinfo
  {journal} {Phys. Rev. D}\ }\textbf {\bibinfo {volume} {99}},\ \bibinfo
  {pages} {023501} (\bibinfo {year} {2019})},\ \Eprint
  {http://arxiv.org/abs/1811.01991} {arXiv:1811.01991 [astro-ph.CO]}
  \BibitemShut {NoStop}%
\bibitem [{\citenamefont {Lorenz}\ \emph {et~al.}(2021)\citenamefont {Lorenz},
  \citenamefont {Funcke}, \citenamefont {L\"offler},\ and\ \citenamefont
  {Calabrese}}]{Lorenz:2021alz}%
  \BibitemOpen
  \bibfield  {author} {\bibinfo {author} {\bibfnamefont {C.~S.}\ \bibnamefont
  {Lorenz}}, \bibinfo {author} {\bibfnamefont {L.}~\bibnamefont {Funcke}},
  \bibinfo {author} {\bibfnamefont {M.}~\bibnamefont {L\"offler}}, \ and\
  \bibinfo {author} {\bibfnamefont {E.}~\bibnamefont {Calabrese}},\ }\href
  {\doibase 10.1103/PhysRevD.104.123518} {\bibfield  {journal} {\bibinfo
  {journal} {Phys. Rev. D}\ }\textbf {\bibinfo {volume} {104}},\ \bibinfo
  {pages} {123518} (\bibinfo {year} {2021})},\ \Eprint
  {http://arxiv.org/abs/2102.13618} {arXiv:2102.13618 [astro-ph.CO]}
  \BibitemShut {NoStop}%
\bibitem [{\citenamefont {Esteban}\ and\ \citenamefont
  {Salvado}(2021)}]{Esteban:2021ozz}%
  \BibitemOpen
  \bibfield  {author} {\bibinfo {author} {\bibfnamefont {I.}~\bibnamefont
  {Esteban}}\ and\ \bibinfo {author} {\bibfnamefont {J.}~\bibnamefont
  {Salvado}},\ }\href {\doibase 10.1088/1475-7516/2021/05/036} {\bibfield
  {journal} {\bibinfo  {journal} {JCAP}\ }\textbf {\bibinfo {volume} {05}},\
  \bibinfo {pages} {036} (\bibinfo {year} {2021})},\ \Eprint
  {http://arxiv.org/abs/2101.05804} {arXiv:2101.05804 [hep-ph]} \BibitemShut
  {NoStop}%
\bibitem [{\citenamefont {Farzan}\ and\ \citenamefont
  {Hannestad}(2016)}]{Farzan:2015pca}%
  \BibitemOpen
  \bibfield  {author} {\bibinfo {author} {\bibfnamefont {Y.}~\bibnamefont
  {Farzan}}\ and\ \bibinfo {author} {\bibfnamefont {S.}~\bibnamefont
  {Hannestad}},\ }\href {\doibase 10.1088/1475-7516/2016/02/058} {\bibfield
  {journal} {\bibinfo  {journal} {JCAP}\ }\textbf {\bibinfo {volume} {02}},\
  \bibinfo {pages} {058} (\bibinfo {year} {2016})},\ \Eprint
  {http://arxiv.org/abs/1510.02201} {arXiv:1510.02201 [hep-ph]} \BibitemShut
  {NoStop}%
\bibitem [{\citenamefont {Renk}\ \emph {et~al.}(2021)\citenamefont {Renk} \emph
  {et~al.}}]{GAMBITCosmologyWorkgroup:2020htv}%
  \BibitemOpen
  \bibfield  {author} {\bibinfo {author} {\bibfnamefont {J.~J.}\ \bibnamefont
  {Renk}} \emph {et~al.} (\bibinfo {collaboration} {GAMBIT Cosmology
  Workgroup}),\ }\href {\doibase 10.1088/1475-7516/2021/02/022} {\bibfield
  {journal} {\bibinfo  {journal} {JCAP}\ }\textbf {\bibinfo {volume} {02}},\
  \bibinfo {pages} {022} (\bibinfo {year} {2021})},\ \Eprint
  {http://arxiv.org/abs/2009.03286} {arXiv:2009.03286 [astro-ph.CO]}
  \BibitemShut {NoStop}%
\bibitem [{\citenamefont {Cuoco}\ \emph {et~al.}(2005)\citenamefont {Cuoco},
  \citenamefont {Lesgourgues}, \citenamefont {Mangano},\ and\ \citenamefont
  {Pastor}}]{Cuoco:2005qr}%
  \BibitemOpen
  \bibfield  {author} {\bibinfo {author} {\bibfnamefont {A.}~\bibnamefont
  {Cuoco}}, \bibinfo {author} {\bibfnamefont {J.}~\bibnamefont {Lesgourgues}},
  \bibinfo {author} {\bibfnamefont {G.}~\bibnamefont {Mangano}}, \ and\
  \bibinfo {author} {\bibfnamefont {S.}~\bibnamefont {Pastor}},\ }\href
  {\doibase 10.1103/PhysRevD.71.123501} {\bibfield  {journal} {\bibinfo
  {journal} {Phys. Rev. D}\ }\textbf {\bibinfo {volume} {71}},\ \bibinfo
  {pages} {123501} (\bibinfo {year} {2005})},\ \Eprint
  {http://arxiv.org/abs/astro-ph/0502465} {arXiv:astro-ph/0502465} \BibitemShut
  {NoStop}%
\bibitem [{\citenamefont {Oldengott}\ \emph {et~al.}(2019)\citenamefont
  {Oldengott}, \citenamefont {Barenboim}, \citenamefont {Kahlen}, \citenamefont
  {Salvado},\ and\ \citenamefont {Schwarz}}]{Oldengott:2019lke}%
  \BibitemOpen
  \bibfield  {author} {\bibinfo {author} {\bibfnamefont {I.~M.}\ \bibnamefont
  {Oldengott}}, \bibinfo {author} {\bibfnamefont {G.}~\bibnamefont
  {Barenboim}}, \bibinfo {author} {\bibfnamefont {S.}~\bibnamefont {Kahlen}},
  \bibinfo {author} {\bibfnamefont {J.}~\bibnamefont {Salvado}}, \ and\
  \bibinfo {author} {\bibfnamefont {D.~J.}\ \bibnamefont {Schwarz}},\ }\href
  {\doibase 10.1088/1475-7516/2019/04/049} {\bibfield  {journal} {\bibinfo
  {journal} {JCAP}\ }\textbf {\bibinfo {volume} {04}},\ \bibinfo {pages} {049}
  (\bibinfo {year} {2019})},\ \Eprint {http://arxiv.org/abs/1901.04352}
  {arXiv:1901.04352 [astro-ph.CO]} \BibitemShut {NoStop}%
\bibitem [{\citenamefont {Akita}\ \emph
  {et~al.}(2021{\natexlab{b}})\citenamefont {Akita}, \citenamefont {Lambiase},\
  and\ \citenamefont {Yamaguchi}}]{Akita:2021hqn}%
  \BibitemOpen
  \bibfield  {author} {\bibinfo {author} {\bibfnamefont {K.}~\bibnamefont
  {Akita}}, \bibinfo {author} {\bibfnamefont {G.}~\bibnamefont {Lambiase}}, \
  and\ \bibinfo {author} {\bibfnamefont {M.}~\bibnamefont {Yamaguchi}},\
  }\href@noop {} {\  (\bibinfo {year} {2021}{\natexlab{b}})},\ \Eprint
  {http://arxiv.org/abs/2109.02900} {arXiv:2109.02900 [hep-ph]} \BibitemShut
  {NoStop}%
\bibitem [{\citenamefont {Ayaita}\ \emph {et~al.}(2016)\citenamefont {Ayaita},
  \citenamefont {Baldi}, \citenamefont {F\"uhrer}, \citenamefont {Puchwein},\
  and\ \citenamefont {Wetterich}}]{Ayaita:2014una}%
  \BibitemOpen
  \bibfield  {author} {\bibinfo {author} {\bibfnamefont {Y.}~\bibnamefont
  {Ayaita}}, \bibinfo {author} {\bibfnamefont {M.}~\bibnamefont {Baldi}},
  \bibinfo {author} {\bibfnamefont {F.}~\bibnamefont {F\"uhrer}}, \bibinfo
  {author} {\bibfnamefont {E.}~\bibnamefont {Puchwein}}, \ and\ \bibinfo
  {author} {\bibfnamefont {C.}~\bibnamefont {Wetterich}},\ }\href {\doibase
  10.1103/PhysRevD.93.063511} {\bibfield  {journal} {\bibinfo  {journal} {Phys.
  Rev. D}\ }\textbf {\bibinfo {volume} {93}},\ \bibinfo {pages} {063511}
  (\bibinfo {year} {2016})},\ \Eprint {http://arxiv.org/abs/1407.8414}
  {arXiv:1407.8414 [astro-ph.CO]} \BibitemShut {NoStop}%
\bibitem [{\citenamefont {Casas}\ \emph {et~al.}(2016)\citenamefont {Casas},
  \citenamefont {Pettorino},\ and\ \citenamefont {Wetterich}}]{Casas:2016duf}%
  \BibitemOpen
  \bibfield  {author} {\bibinfo {author} {\bibfnamefont {S.}~\bibnamefont
  {Casas}}, \bibinfo {author} {\bibfnamefont {V.}~\bibnamefont {Pettorino}}, \
  and\ \bibinfo {author} {\bibfnamefont {C.}~\bibnamefont {Wetterich}},\ }\href
  {\doibase 10.1103/PhysRevD.94.103518} {\bibfield  {journal} {\bibinfo
  {journal} {Phys. Rev. D}\ }\textbf {\bibinfo {volume} {94}},\ \bibinfo
  {pages} {103518} (\bibinfo {year} {2016})},\ \Eprint
  {http://arxiv.org/abs/1608.02358} {arXiv:1608.02358 [astro-ph.CO]}
  \BibitemShut {NoStop}%
\bibitem [{\citenamefont {Sabti}\ \emph {et~al.}(2020)\citenamefont {Sabti},
  \citenamefont {Magalich},\ and\ \citenamefont {Filimonova}}]{Sabti:2020yrt}%
  \BibitemOpen
  \bibfield  {author} {\bibinfo {author} {\bibfnamefont {N.}~\bibnamefont
  {Sabti}}, \bibinfo {author} {\bibfnamefont {A.}~\bibnamefont {Magalich}}, \
  and\ \bibinfo {author} {\bibfnamefont {A.}~\bibnamefont {Filimonova}},\
  }\href {\doibase 10.1088/1475-7516/2020/11/056} {\bibfield  {journal}
  {\bibinfo  {journal} {JCAP}\ }\textbf {\bibinfo {volume} {11}},\ \bibinfo
  {pages} {056} (\bibinfo {year} {2020})},\ \Eprint
  {http://arxiv.org/abs/2006.07387} {arXiv:2006.07387 [hep-ph]} \BibitemShut
  {NoStop}%
\bibitem [{\citenamefont {Ringwald}\ and\ \citenamefont
  {Wong}(2004)}]{Ringwald:2004np}%
  \BibitemOpen
  \bibfield  {author} {\bibinfo {author} {\bibfnamefont {A.}~\bibnamefont
  {Ringwald}}\ and\ \bibinfo {author} {\bibfnamefont {Y.~Y.~Y.}\ \bibnamefont
  {Wong}},\ }\href {\doibase 10.1088/1475-7516/2004/12/005} {\bibfield
  {journal} {\bibinfo  {journal} {JCAP}\ }\textbf {\bibinfo {volume} {12}},\
  \bibinfo {pages} {005} (\bibinfo {year} {2004})},\ \Eprint
  {http://arxiv.org/abs/hep-ph/0408241} {arXiv:hep-ph/0408241} \BibitemShut
  {NoStop}%
\bibitem [{\citenamefont {de~Salas}\ \emph {et~al.}(2017)\citenamefont
  {de~Salas}, \citenamefont {Gariazzo}, \citenamefont {Lesgourgues},\ and\
  \citenamefont {Pastor}}]{deSalas:2017wtt}%
  \BibitemOpen
  \bibfield  {author} {\bibinfo {author} {\bibfnamefont {P.~F.}\ \bibnamefont
  {de~Salas}}, \bibinfo {author} {\bibfnamefont {S.}~\bibnamefont {Gariazzo}},
  \bibinfo {author} {\bibfnamefont {J.}~\bibnamefont {Lesgourgues}}, \ and\
  \bibinfo {author} {\bibfnamefont {S.}~\bibnamefont {Pastor}},\ }\href
  {\doibase 10.1088/1475-7516/2017/09/034} {\bibfield  {journal} {\bibinfo
  {journal} {JCAP}\ }\textbf {\bibinfo {volume} {09}},\ \bibinfo {pages} {034}
  (\bibinfo {year} {2017})},\ \Eprint {http://arxiv.org/abs/1706.09850}
  {arXiv:1706.09850 [astro-ph.CO]} \BibitemShut {NoStop}%
\bibitem [{\citenamefont {Zhang}\ and\ \citenamefont
  {Zhang}(2018)}]{Zhang:2017ljh}%
  \BibitemOpen
  \bibfield  {author} {\bibinfo {author} {\bibfnamefont {J.}~\bibnamefont
  {Zhang}}\ and\ \bibinfo {author} {\bibfnamefont {X.}~\bibnamefont {Zhang}},\
  }\href {\doibase 10.1038/s41467-018-04264-y} {\bibfield  {journal} {\bibinfo
  {journal} {Nature Commun.}\ }\textbf {\bibinfo {volume} {9}},\ \bibinfo
  {pages} {1833} (\bibinfo {year} {2018})},\ \Eprint
  {http://arxiv.org/abs/1712.01153} {arXiv:1712.01153 [astro-ph.CO]}
  \BibitemShut {NoStop}%
\bibitem [{\citenamefont {Mertsch}\ \emph {et~al.}(2020)\citenamefont
  {Mertsch}, \citenamefont {Parimbelli}, \citenamefont {de~Salas},
  \citenamefont {Gariazzo}, \citenamefont {Lesgourgues},\ and\ \citenamefont
  {Pastor}}]{Mertsch:2019qjv}%
  \BibitemOpen
  \bibfield  {author} {\bibinfo {author} {\bibfnamefont {P.}~\bibnamefont
  {Mertsch}}, \bibinfo {author} {\bibfnamefont {G.}~\bibnamefont {Parimbelli}},
  \bibinfo {author} {\bibfnamefont {P.~F.}\ \bibnamefont {de~Salas}}, \bibinfo
  {author} {\bibfnamefont {S.}~\bibnamefont {Gariazzo}}, \bibinfo {author}
  {\bibfnamefont {J.}~\bibnamefont {Lesgourgues}}, \ and\ \bibinfo {author}
  {\bibfnamefont {S.}~\bibnamefont {Pastor}},\ }\href {\doibase
  10.1088/1475-7516/2020/01/015} {\bibfield  {journal} {\bibinfo  {journal}
  {JCAP}\ }\textbf {\bibinfo {volume} {01}},\ \bibinfo {pages} {015} (\bibinfo
  {year} {2020})},\ \Eprint {http://arxiv.org/abs/1910.13388} {arXiv:1910.13388
  [astro-ph.CO]} \BibitemShut {NoStop}%
\bibitem [{\citenamefont {Piffl}\ \emph {et~al.}(2014)\citenamefont {Piffl}
  \emph {et~al.}}]{Piffl:2013mla}%
  \BibitemOpen
  \bibfield  {author} {\bibinfo {author} {\bibfnamefont {T.}~\bibnamefont
  {Piffl}} \emph {et~al.},\ }\href {\doibase 10.1051/0004-6361/201322531}
  {\bibfield  {journal} {\bibinfo  {journal} {Astron. Astrophys.}\ }\textbf
  {\bibinfo {volume} {562}},\ \bibinfo {pages} {A91} (\bibinfo {year}
  {2014})},\ \Eprint {http://arxiv.org/abs/1309.4293} {arXiv:1309.4293
  [astro-ph.GA]} \BibitemShut {NoStop}%
\bibitem [{\citenamefont {Roulet}\ and\ \citenamefont
  {Vissani}(2018)}]{Roulet:2018fyh}%
  \BibitemOpen
  \bibfield  {author} {\bibinfo {author} {\bibfnamefont {E.}~\bibnamefont
  {Roulet}}\ and\ \bibinfo {author} {\bibfnamefont {F.}~\bibnamefont
  {Vissani}},\ }\href {\doibase 10.1088/1475-7516/2018/10/049} {\bibfield
  {journal} {\bibinfo  {journal} {JCAP}\ }\textbf {\bibinfo {volume} {10}},\
  \bibinfo {pages} {049} (\bibinfo {year} {2018})},\ \Eprint
  {http://arxiv.org/abs/1810.00505} {arXiv:1810.00505 [hep-ph]} \BibitemShut
  {NoStop}%
\bibitem [{\citenamefont {Ade}\ \emph {et~al.}(2019)\citenamefont {Ade} \emph
  {et~al.}}]{SimonsObservatory:2018koc}%
  \BibitemOpen
  \bibfield  {author} {\bibinfo {author} {\bibfnamefont {P.}~\bibnamefont
  {Ade}} \emph {et~al.} (\bibinfo {collaboration} {Simons Observatory}),\
  }\href {\doibase 10.1088/1475-7516/2019/02/056} {\bibfield  {journal}
  {\bibinfo  {journal} {JCAP}\ }\textbf {\bibinfo {volume} {02}},\ \bibinfo
  {pages} {056} (\bibinfo {year} {2019})},\ \Eprint
  {http://arxiv.org/abs/1808.07445} {arXiv:1808.07445 [astro-ph.CO]}
  \BibitemShut {NoStop}%
\bibitem [{\citenamefont {Amendola}\ \emph {et~al.}(2008)\citenamefont
  {Amendola}, \citenamefont {Baldi},\ and\ \citenamefont
  {Wetterich}}]{Amendola:2007yx}%
  \BibitemOpen
  \bibfield  {author} {\bibinfo {author} {\bibfnamefont {L.}~\bibnamefont
  {Amendola}}, \bibinfo {author} {\bibfnamefont {M.}~\bibnamefont {Baldi}}, \
  and\ \bibinfo {author} {\bibfnamefont {C.}~\bibnamefont {Wetterich}},\ }\href
  {\doibase 10.1103/PhysRevD.78.023015} {\bibfield  {journal} {\bibinfo
  {journal} {Phys. Rev. D}\ }\textbf {\bibinfo {volume} {78}},\ \bibinfo
  {pages} {023015} (\bibinfo {year} {2008})},\ \Eprint
  {http://arxiv.org/abs/0706.3064} {arXiv:0706.3064 [astro-ph]} \BibitemShut
  {NoStop}%
\bibitem [{\citenamefont {Wetterich}(2007)}]{Wetterich:2007kr}%
  \BibitemOpen
  \bibfield  {author} {\bibinfo {author} {\bibfnamefont {C.}~\bibnamefont
  {Wetterich}},\ }\href {\doibase 10.1016/j.physletb.2007.08.060} {\bibfield
  {journal} {\bibinfo  {journal} {Phys. Lett. B}\ }\textbf {\bibinfo {volume}
  {655}},\ \bibinfo {pages} {201} (\bibinfo {year} {2007})},\ \Eprint
  {http://arxiv.org/abs/0706.4427} {arXiv:0706.4427 [hep-ph]} \BibitemShut
  {NoStop}%
\bibitem [{\citenamefont {Pettorino}\ \emph {et~al.}(2010)\citenamefont
  {Pettorino}, \citenamefont {Wintergerst}, \citenamefont {Amendola},\ and\
  \citenamefont {Wetterich}}]{Pettorino:2010bv}%
  \BibitemOpen
  \bibfield  {author} {\bibinfo {author} {\bibfnamefont {V.}~\bibnamefont
  {Pettorino}}, \bibinfo {author} {\bibfnamefont {N.}~\bibnamefont
  {Wintergerst}}, \bibinfo {author} {\bibfnamefont {L.}~\bibnamefont
  {Amendola}}, \ and\ \bibinfo {author} {\bibfnamefont {C.}~\bibnamefont
  {Wetterich}},\ }\href {\doibase 10.1103/PhysRevD.82.123001} {\bibfield
  {journal} {\bibinfo  {journal} {Phys. Rev. D}\ }\textbf {\bibinfo {volume}
  {82}},\ \bibinfo {pages} {123001} (\bibinfo {year} {2010})},\ \Eprint
  {http://arxiv.org/abs/1009.2461} {arXiv:1009.2461 [astro-ph.CO]} \BibitemShut
  {NoStop}%
\bibitem [{\citenamefont {Roy~Choudhury}\ and\ \citenamefont
  {Hannestad}(2020)}]{RoyChoudhury:2019hls}%
  \BibitemOpen
  \bibfield  {author} {\bibinfo {author} {\bibfnamefont {S.}~\bibnamefont
  {Roy~Choudhury}}\ and\ \bibinfo {author} {\bibfnamefont {S.}~\bibnamefont
  {Hannestad}},\ }\href {\doibase 10.1088/1475-7516/2020/07/037} {\bibfield
  {journal} {\bibinfo  {journal} {JCAP}\ }\textbf {\bibinfo {volume} {07}},\
  \bibinfo {pages} {037} (\bibinfo {year} {2020})},\ \Eprint
  {http://arxiv.org/abs/1907.12598} {arXiv:1907.12598 [astro-ph.CO]}
  \BibitemShut {NoStop}%
\bibitem [{\citenamefont {Di~Valentino}\ \emph {et~al.}(2020)\citenamefont
  {Di~Valentino}, \citenamefont {Melchiorri},\ and\ \citenamefont
  {Silk}}]{DiValentino:2019dzu}%
  \BibitemOpen
  \bibfield  {author} {\bibinfo {author} {\bibfnamefont {E.}~\bibnamefont
  {Di~Valentino}}, \bibinfo {author} {\bibfnamefont {A.}~\bibnamefont
  {Melchiorri}}, \ and\ \bibinfo {author} {\bibfnamefont {J.}~\bibnamefont
  {Silk}},\ }\href {\doibase 10.1088/1475-7516/2020/01/013} {\bibfield
  {journal} {\bibinfo  {journal} {JCAP}\ }\textbf {\bibinfo {volume} {01}},\
  \bibinfo {pages} {013} (\bibinfo {year} {2020})},\ \Eprint
  {http://arxiv.org/abs/1908.01391} {arXiv:1908.01391 [astro-ph.CO]}
  \BibitemShut {NoStop}%
\bibitem [{\citenamefont {Fardon}\ \emph {et~al.}(2004)\citenamefont {Fardon},
  \citenamefont {Nelson},\ and\ \citenamefont {Weiner}}]{Fardon:2003eh}%
  \BibitemOpen
  \bibfield  {author} {\bibinfo {author} {\bibfnamefont {R.}~\bibnamefont
  {Fardon}}, \bibinfo {author} {\bibfnamefont {A.~E.}\ \bibnamefont {Nelson}},
  \ and\ \bibinfo {author} {\bibfnamefont {N.}~\bibnamefont {Weiner}},\ }\href
  {\doibase 10.1088/1475-7516/2004/10/005} {\bibfield  {journal} {\bibinfo
  {journal} {JCAP}\ }\textbf {\bibinfo {volume} {10}},\ \bibinfo {pages} {005}
  (\bibinfo {year} {2004})},\ \Eprint {http://arxiv.org/abs/astro-ph/0309800}
  {arXiv:astro-ph/0309800} \BibitemShut {NoStop}%
\bibitem [{\citenamefont {Peccei}(2005)}]{Peccei:2004sz}%
  \BibitemOpen
  \bibfield  {author} {\bibinfo {author} {\bibfnamefont {R.~D.}\ \bibnamefont
  {Peccei}},\ }\href {\doibase 10.1103/PhysRevD.71.023527} {\bibfield
  {journal} {\bibinfo  {journal} {Phys. Rev. D}\ }\textbf {\bibinfo {volume}
  {71}},\ \bibinfo {pages} {023527} (\bibinfo {year} {2005})},\ \Eprint
  {http://arxiv.org/abs/hep-ph/0411137} {arXiv:hep-ph/0411137} \BibitemShut
  {NoStop}%
\bibitem [{\citenamefont {Afshordi}\ \emph {et~al.}(2005)\citenamefont
  {Afshordi}, \citenamefont {Zaldarriaga},\ and\ \citenamefont
  {Kohri}}]{Afshordi:2005ym}%
  \BibitemOpen
  \bibfield  {author} {\bibinfo {author} {\bibfnamefont {N.}~\bibnamefont
  {Afshordi}}, \bibinfo {author} {\bibfnamefont {M.}~\bibnamefont
  {Zaldarriaga}}, \ and\ \bibinfo {author} {\bibfnamefont {K.}~\bibnamefont
  {Kohri}},\ }\href {\doibase 10.1103/PhysRevD.72.065024} {\bibfield  {journal}
  {\bibinfo  {journal} {Phys. Rev. D}\ }\textbf {\bibinfo {volume} {72}},\
  \bibinfo {pages} {065024} (\bibinfo {year} {2005})},\ \Eprint
  {http://arxiv.org/abs/astro-ph/0506663} {arXiv:astro-ph/0506663} \BibitemShut
  {NoStop}%
\bibitem [{\citenamefont {Smirnov}\ and\ \citenamefont
  {Xu}(2022)}]{Smirnov:2022sfo}%
  \BibitemOpen
  \bibfield  {author} {\bibinfo {author} {\bibfnamefont {A.~Y.}\ \bibnamefont
  {Smirnov}}\ and\ \bibinfo {author} {\bibfnamefont {X.-J.}\ \bibnamefont
  {Xu}},\ }\href@noop {} {\  (\bibinfo {year} {2022})},\ \Eprint
  {http://arxiv.org/abs/2201.00939} {arXiv:2201.00939 [hep-ph]} \BibitemShut
  {NoStop}%
\bibitem [{\citenamefont {Mikulenko}\ \emph {et~al.}(2021)\citenamefont
  {Mikulenko}, \citenamefont {Cheipesh}, \citenamefont {Cheianov},\ and\
  \citenamefont {Boyarsky}}]{Mikulenko:2021ydo}%
  \BibitemOpen
  \bibfield  {author} {\bibinfo {author} {\bibfnamefont {O.}~\bibnamefont
  {Mikulenko}}, \bibinfo {author} {\bibfnamefont {Y.}~\bibnamefont {Cheipesh}},
  \bibinfo {author} {\bibfnamefont {V.}~\bibnamefont {Cheianov}}, \ and\
  \bibinfo {author} {\bibfnamefont {A.}~\bibnamefont {Boyarsky}},\ }\href@noop
  {} {\  (\bibinfo {year} {2021})},\ \Eprint {http://arxiv.org/abs/2111.09292}
  {arXiv:2111.09292 [hep-ph]} \BibitemShut {NoStop}%
\bibitem [{\citenamefont {Brdar}\ \emph {et~al.}(2022)\citenamefont {Brdar},
  \citenamefont {Plestid},\ and\ \citenamefont {Rocco}}]{Brdar:2022wuv}%
  \BibitemOpen
  \bibfield  {author} {\bibinfo {author} {\bibfnamefont {V.}~\bibnamefont
  {Brdar}}, \bibinfo {author} {\bibfnamefont {R.}~\bibnamefont {Plestid}}, \
  and\ \bibinfo {author} {\bibfnamefont {N.}~\bibnamefont {Rocco}},\
  }\href@noop {} {\  (\bibinfo {year} {2022})},\ \Eprint
  {http://arxiv.org/abs/2201.07251} {arXiv:2201.07251 [hep-ph]} \BibitemShut
  {NoStop}%
\bibitem [{\citenamefont {Lesgourgues}(2011)}]{Lesgourgues:2011re}%
  \BibitemOpen
  \bibfield  {author} {\bibinfo {author} {\bibfnamefont {J.}~\bibnamefont
  {Lesgourgues}},\ }\href@noop {} {\  (\bibinfo {year} {2011})},\ \Eprint
  {http://arxiv.org/abs/1104.2932} {arXiv:1104.2932 [astro-ph.IM]} \BibitemShut
  {NoStop}%
\bibitem [{\citenamefont {Blas}\ \emph {et~al.}(2011)\citenamefont {Blas},
  \citenamefont {Lesgourgues},\ and\ \citenamefont {Tram}}]{Blas:2011rf}%
  \BibitemOpen
  \bibfield  {author} {\bibinfo {author} {\bibfnamefont {D.}~\bibnamefont
  {Blas}}, \bibinfo {author} {\bibfnamefont {J.}~\bibnamefont {Lesgourgues}}, \
  and\ \bibinfo {author} {\bibfnamefont {T.}~\bibnamefont {Tram}},\ }\href
  {\doibase 10.1088/1475-7516/2011/07/034} {\bibfield  {journal} {\bibinfo
  {journal} {JCAP}\ }\textbf {\bibinfo {volume} {07}},\ \bibinfo {pages} {034}
  (\bibinfo {year} {2011})},\ \Eprint {http://arxiv.org/abs/1104.2933}
  {arXiv:1104.2933 [astro-ph.CO]} \BibitemShut {NoStop}%
\end{thebibliography}%

\end{document}